%% file: main.tex
\documentclass[lettersize,journal]{IEEEtran}
\usepackage{amsmath,amsfonts}
\usepackage{algorithmic}
\usepackage{algorithm}
\usepackage{array}
\usepackage[caption=false,font=normalsize,labelfont=sf,textfont=sf]{subfig}
\usepackage{textcomp}
\usepackage{stfloats}
\usepackage{url}
\usepackage{verbatim}
\usepackage{graphicx}
\usepackage{cite}

\usepackage{bm}
\usepackage{pifont}

\usepackage{colortbl}

\usepackage{multirow}

\usepackage{xcolor}

\begin{document}

\title{Joint Beam Search Integrating CTC, Attention, \\and Transducer Decoders}

\author{Yui Sudo,~\IEEEmembership{Member,~IEEE,} 
        Muhammad Shakeel,~\IEEEmembership{Non-Member,~IEEE,} 
        Yosuke Fukumoto,~\IEEEmembership{Non-Member,}\\
        Brian Yan,~\IEEEmembership{Student Member,~IEEE,} 
        Jiatong Shi,~\IEEEmembership{Student Member,~IEEE,} 
        Yifan Peng,~\IEEEmembership{Student Member,~IEEE,} 
        Shinji Watanabe,~\IEEEmembership{Fellow,~IEEE,}
\thanks{This paper was produced by the IEEE Publication Technology Group. They are in Piscataway, NJ.}
\thanks{Manuscript received May 23, 2024.}}

\markboth{Journal of \LaTeX\ Class Files,~Vol.~14, No.~8, August~2024}%
{Shell \MakeLowercase{\textit{et al.}}: A Sample Article Using IEEEtran.cls for IEEE Journals}

\IEEEpubid{0000--0000/00\$00.00~\copyright~2024 IEEE}

\maketitle

\begin{abstract}
End-to-end automatic speech recognition (E2E-ASR) can be classified by its decoder architectures, such as connectionist temporal classification (CTC), recurrent neural network transducer (RNN-T), attention-based encoder-decoder, and Mask-CTC models. 
Each decoder architecture has advantages and disadvantages, leading practitioners to switch between these different models depending on application requirements.
Instead of building separate models, we propose a joint modeling scheme where four decoders (CTC, RNN-T, attention, and Mask-CTC) share the same encoder -- we refer to this as 4D modeling. The 4D model is trained jointly, which will bring model regularization and maximize the model robustness thanks to their complementary properties.
To efficiently train the 4D model, we introduce a two-stage training strategy that stabilizes the joint training.
In addition, we propose three novel joint beam search algorithms by combining three decoders (CTC, RNN-T, and attention) to further improve performance.
These three beam search algorithms differ in which decoder is used as the primary decoder. We carefully evaluate the performance and computational tradeoffs associated with each algorithm. 
Experimental results demonstrate that the jointly trained 4D model outperforms the E2E-ASR models trained with only one individual decoder.
Furthermore, we demonstrate that the proposed joint beam search algorithm outperforms the previously proposed CTC/attention decoding.
\end{abstract}

\begin{IEEEkeywords}
Speech recognition, CTC, attention-based encoder-decoder, RNN-T, beam search
\end{IEEEkeywords}

\section{Introduction}
\label{sec:intro}

End-to-end automatic speech recognition (E2E-ASR) \cite{prabhavalkar2023end,li2022recent} has been a subject of active research. 
Existing E2E-ASR systems can be classified by their decoder architectures: connectionist temporal classification (CTC) \cite{ctc1,ctc2,kriman2020quartznet}, recurrent neural network transducer (RNN-T) \cite{rnnt1,zhang2020transformer,han2020contextnet,rnnt2}, attention-based encoder-decoder \cite{attention1,attention2,karita2019comparative,guo2021recent}, and non-autoregressive (NAR) methods \cite{higuchi2020mask,chen2020non,song2021non}.
These decoders align speech signals and token sequences in various ways, each with different strengths and weaknesses, as follows:
\begin{itemize}
\vspace*{-0mm}
\leftskip -2.0mm 
    \item CTC predicts a monotonic alignment of output tokens with input speech frames efficiently with the conditional independence assumption. While its parallelized inference across frames is fast, the conditional independence assumption results in suboptimal performance. CTC is also valuable for segmenting long recordings \cite{ctcsegmentation}.
    \item RNN-T shares a monotonic alignment property with CTC but relaxes the conditional independence assumption. It typically outperforms CTC thanks to this relaxation and is particularly adept for streaming ASR \cite{rnnt2}. However, due to its flexible alignment paths, its larger modeling space usually leads to additional training difficulties \cite{kuang22_interspeech}.
    \item Attention-based encoder-decoder employs a cross-attention mechanism to enable flexible alignments between input and output sequences, making it useful for various tasks such as translation \cite{bahdanau2014neural,pmlr-v202-radford23a}. However, it may be susceptible to alignment errors in ASR tasks due to the lack of a monotonicity constraint \cite{watanabe2017hybrid}.
    \item NAR methods, as exemplified by Mask-CTC \cite{higuchi2020mask}, estimate the token sequence using the entire input sequence. The mask-based approach \cite{ghazvininejad2019mask} maintains fast parallelized inference across frames, but unlike CTC, it accounts for label dependencies. It is also applicable in two-pass rescoring approaches for ASR error correction \cite{futami2022correction}.
\end{itemize}

\IEEEpubidadjcol

Owing to distinct characteristics in each decoder, it is common to use corresponding models according to different application scenarios. For example, CTC is suitable for on-device systems with lower computational demands, whereas attention-based encoder-decoder is preferred for offline systems with less stringent latency requirements. However, maintaining multiple models for different application scenarios creates additional maintenance overhead for system administrators. 
Thus, several initiatives have attempted to integrate these models, such as hybrid CTC/attention, to mitigate their shortcomings \cite{watanabe2017hybrid,ueno2018acoustic,nakatani2019improving}. Joint training in hybrid CTC/attention, achieved by sharing an encoder, helps to regularize the over-flexibility of the attention mechanism and the conditional independence assumption of the CTC model. During decoding, a joint beam search involving both CTC and attention decoders is employed to enhance performance further \cite{watanabe2017hybrid}. 
Such joint modeling has also been applied to large-scale E2E-ASR models \cite{peng2023reproducing,peng2024owsm,zhang2022wenetspeech,fujimotoreazonspeech}. 
Other integrated models, utilizing two-pass decoding of RNN-T/attention, NAR/attention, and RNN-T/NAR, have also been proposed \cite{sainath2019two,hu2020deliberation,hu2021transformer,tian2022hybrid,wang2022deliberation,yao2021wenet}. These methods attempt to combine a streaming decoder with an offline decoder to achieve both low latency and high accuracy.
In addition, several approaches have been explored to combine dual RNN-T models with different context lengths to optimize both latency and accuracy in ASR systems \cite{Narayanan2020CascadedEF,mahadeokar2022streaming,li2022improving,yu2021dualmode,moritz2021dual,weninger2022conformer}.
While integrating two of these models has shown effectiveness, a more comprehensive integration of multiple models can exploit their complementary features to a greater extent. A single model can realize broader applications by minimizing the overhead of maintaining separate models.

In this study, we seek to jointly model four decoders (4D) with a shared encoder: 
CTC, attention, RNN-T, and Mask-CTC. 
The 4D model is trained jointly, which will bring model regularization and maximize the model robustness thanks to their complementary properties.
The jointly trained 4D model allows the selection of an appropriate decoder depending on different application scenarios. 
In addition, we propose three novel joint beam search algorithms using CTC, RNN-T, and attention to further enhance the complementarity of each decoder during inference. 
Note that the Mask-CTC is excluded from the proposed beam search because it predicts the entire output sequence in parallel, unlike the other three decoders.

The remainder of the manuscript is organized as follows: 
Section~\ref{sec:related} describes the related work with a focus on the joint modeling of multiple decoders, highlighting the main contributions of this paper. 
Section~\ref{sec:Preliminary} provides an overview of the four decoders, which are extended to the proposed 4D model in Section~\ref{sec:proposed}. 
Sections~\ref{sec:experimental condition} and \ref{sec:experiments} describe the experimental setup and results. Section~\ref{sec:discussion} discusses the practicality of the proposed method based on the experimental results, and Section~\ref{sec:conclusion} concludes the paper.

\section{Related work}
\label{sec:related}

This section describes the joint modeling approach, 
\textcolor{black}{
ranging from deep neural network--hidden Markov model (DNN-HMM) \cite{hinton2012deep,seide11_interspeech} to E2E-ASR systems.
}

\subsection{\textcolor{black}{Joint modeling of DNN-HMM systems}}

\textcolor{black}{
Prior to the advent of E2E-ASR, extensive research has explored joint modeling techniques for DNN-HMM systems \cite{XU2011802,6638967,wang15k_interspeech,7472764}. These methods improve ASR performance by combining scores from multiple systems based on either manually tuned or estimated weights \cite{7472764}.
The joint modeling approaches have been further extended to integrate DNN-HMM systems with E2E-ASR models \cite{wong20_interspeech,alumae21_interspeech,9747144,LI202312}. This body of work has demonstrated the potential of joint modeling to improve ASR performance across diverse system architectures.
}

\subsection{\textcolor{black}{Joint modeling of E2E-ASR systems}}

\textcolor{black}{
Building on these advancements, the joint modeling approaches have been extended to E2E-ASR systems. These approaches} mainly fall into two categories: two-pass rescoring and joint beam search.

\subsubsection{Two-pass rescoring}
In the two-pass rescoring methods, the primary decoder generates complete hypotheses, followed by the other decoder rescoring those hypotheses to produce a more probable token sequence \cite{sainath2019two,hu2020deliberation,hu2021transformer,tian2022hybrid,wang2022deliberation,yao2021wenet}.
For example, in the two-pass rescoring of RNN-T/attention \cite{sainath2019two,hu2020deliberation,hu2021transformer}, the RNN-T decoder generates complete n-best hypotheses, and the attention decoder subsequently rescores them, leading to enhanced performance. However, the effectiveness of two-pass rescoring heavily relies on the first pass, because the second pass cannot correct the hypothesis unless the first pass contains the correct one \cite{watanabe2017hybrid}.

\subsubsection{Joint beam search}
Conversely, the joint beam search approach performs joint rescoring using the secondary decoder during the beam search process, without waiting for the primary decoder to complete hypothesis expansion. 
A typical example is the attention-driven joint beam search based on the hybrid CTC/attention \cite{watanabe2017hybrid}. In this method, the attention decoder serves as the primary decoder, and the CTC decoder scores the hypotheses generated by the attention decoder in a label-synchronous manner. This method integrates the two decoders more tightly, resulting in more accurate decoding than two-pass rescoring.
This method has been extended to a CTC-driven joint beam search algorithm based on the CTC/attention model for speech translation tasks \cite{yan2022ctc}. This method is computationally faster than the attention-driven joint beam search algorithm with almost comparable performance.
Note that while both methods use the same CTC/attention model, the choice of primary decoder affects their performance and computational cost. 
Other method includes the integration of time- and label-synchronous beam search using the streaming CTC/attention and the time-synchronous joint beam search using online/offline RNN-T decoders \cite{tsunoo23_interspeech,sudo23c_interspeech}. 

\subsection{Limitations of the previous work}
While these methods have been successful with combinations of two decoders, there are still limited efforts in integrating three or more decoders. Additionally, only limited combinations of decoders have been considered for joint beam search (e.g., CTC/attention and online/offline RNN-T). 
As a result, the optimal joint modeling strategy remains unclear.
Furthermore, approaches that do not use a joint model require the training of separate models for each application scenario, increasing the complexity of deployment.
Given that different primary decoders result in differences in performance and computational cost \cite{yan2022ctc}, a more comprehensive analysis is required to understand the tradeoffs between different decoder combinations and identify the best approach for different needs.

\subsection{Contributions of this work}
Therefore, this paper proposes the 4D model consisting of CTC, attention, RNN-T, and Mask-CTC with a shared encoder.
The joint training of these four decoders not only leverages their complementary properties for model regularization but also allows for flexibility in choosing the appropriate decoding approach depending on the application constraints and requirements.
Furthermore, we introduce three novel joint beam search algorithms using CTC, RNN-T, and attention to enhance the complementarity of each decoder during inference. 
By thoroughly evaluating the performance and tradeoffs associated with different primary decoders, this paper provides practical insights into optimizing recognition performance while managing computational cost.
Specifically, we
\begin{itemize}
\leftskip -2.5mm 
    \item Demonstrate that each single-decoder of the 4D model is improved over counterparts without joint training with the efficient two-stage optimization strategy.
    \item Introduce novel joint beam search algorithms for joint CTC/RNN-T/attention decoding, showcasing superior performance on average compared to CTC/attention decoding.
    \item Analyze the effect of integrating multiple decoders in the beam search by carefully comparing the three joint beam search algorithms.
\end{itemize}
This paper extends our previous study on the 4D model \cite{sudo20224d} by newly proposing the attention-driven joint beam search for CTC/RNN-T/attention decoders (Section~\ref{sec:proposed}). 
We also analyze the effect of integrating multiple decoders by carefully comparing the three joint beam search algorithms including a more comprehensive set of experiments (Section~\ref{sec:experiments}).
Specifically, in Section \ref{sec:experiments}-A, we provide a detailed error analysis, including the language model (LM) shallow fusion. In Section \ref{sec:experiments}-B, we investigate how the number of decoders used during joint training affects performance and training time. 
In Sections \ref{sec:experiments}-C and \ref{sec:experiments}-D, we further explore the tradeoff between real-time factor (RTF) and word error rate (WER), as well as the computational cost in terms of input/output length and beam size. 
In addition, we provide detailed formulations of the four decoders (Section~\ref{sec:Preliminary}).

\section{Preliminary}
\label{sec:Preliminary}

This section describes the conformer-based encoder, commonly used across the four decoders: CTC, RNN-T, attention, and Mask-CTC. Subsequently, we provide a detailed description of each decoder.

\subsection{Encoder}
\label{sec:encoder}

We adopt the conformer \cite{guo2021recent,gulati2020conformer} as the shared encoder in this study.
The encoder comprises two convolutional layers and $N$ conformer layers.
The convolutional layers subsample a $L$-length audio feature sequence, $X = (\bm{x}_l \in \mathbb{R}^{d} \mid l = 1, ..., L)$ into a $T ( < L)$-length subsampled feature sequence, $U = (\bm{u}_t \in \mathbb{R}^{d^{\prime}} \mid t = 1, ..., T)$, as follows: 
\begin{align}
    U = \mathrm{ConvSubsamp}(X).
\end{align}
Here, $\bm{x}_l$ and $\bm{u}_t$ are a $d$-dimensional speech feature vector (e.g., log Mel filterbanks) and $d^{\prime}$-dimensional hidden state vector, respectively.
Subsequently, the $N$ conformer blocks transform the subsampled feature sequence $U$ into a $T$-length hidden state sequence, $H = (\bm{h}_t \in \mathbb{R}^{d'} \mid t = 1, ..., T)$, described as,
\begin{equation}
\label{eq:conformer-encoder}
    H = \mathrm{Conformer}(U).
\end{equation}
Each conformer block consists of a feed-forward layer (FFN), a multiheaded self-attention layer (MHSA), a convolution layer (Conv), and another FFN module, with layer normalization layers (LN) and residual connections \cite{ba2016layer,7780459} as follows:
\begin{align}
    &\tilde{U_n} = U_n + \frac{1}{2}\mathrm{FFN}(\mathrm{LN}(U_n)),\\
    &U^{\prime}_n = \tilde{U_n} + \mathrm{MHSA}(\mathrm{LN}(\tilde{U_n})),\\
    &U^{\prime\prime}_n = U^{\prime}_n + \mathrm{Conv}(\mathrm{LN}(U^{\prime}_n)),\\
    &U^{\prime\prime\prime}_n = U^{\prime\prime}_n + \frac{1}{2}\mathrm{FFN}(\mathrm{LN}(U^{\prime\prime}_n)),
\end{align}
where $U_n$ and $U^{\prime\prime\prime}_n$ represent the input and output of the $n$-th conformer block, respectively ($U_1 = U$ in Eq.~\eqref{eq:conformer-encoder}).  
Note that the LN layers are applied before each module followed by a residual connection as in \cite{guo2021recent}.
By using $N$ conformer blocks, the conformer encoder outputs the hidden state sequence ($H = U^{\prime\prime\prime}_N$).

The hidden state sequence $H$ extracted by the encoder is fed into the decoder and the decoder estimates the $S$-length output token sequence, $Y = (y_s \in \mathcal{V} \mid s = 1, ..., S)$, where $\mathcal{V}$ and $s$ represent the vocabulary and the label index, respectively.
Each decoder is described in the following sections.

\subsection{CTC}
\label{sec:ctc}

\begin{figure}[t!]
    \centering
        \begin{minipage}{0.49\textwidth}
            \includegraphics[width=\textwidth]{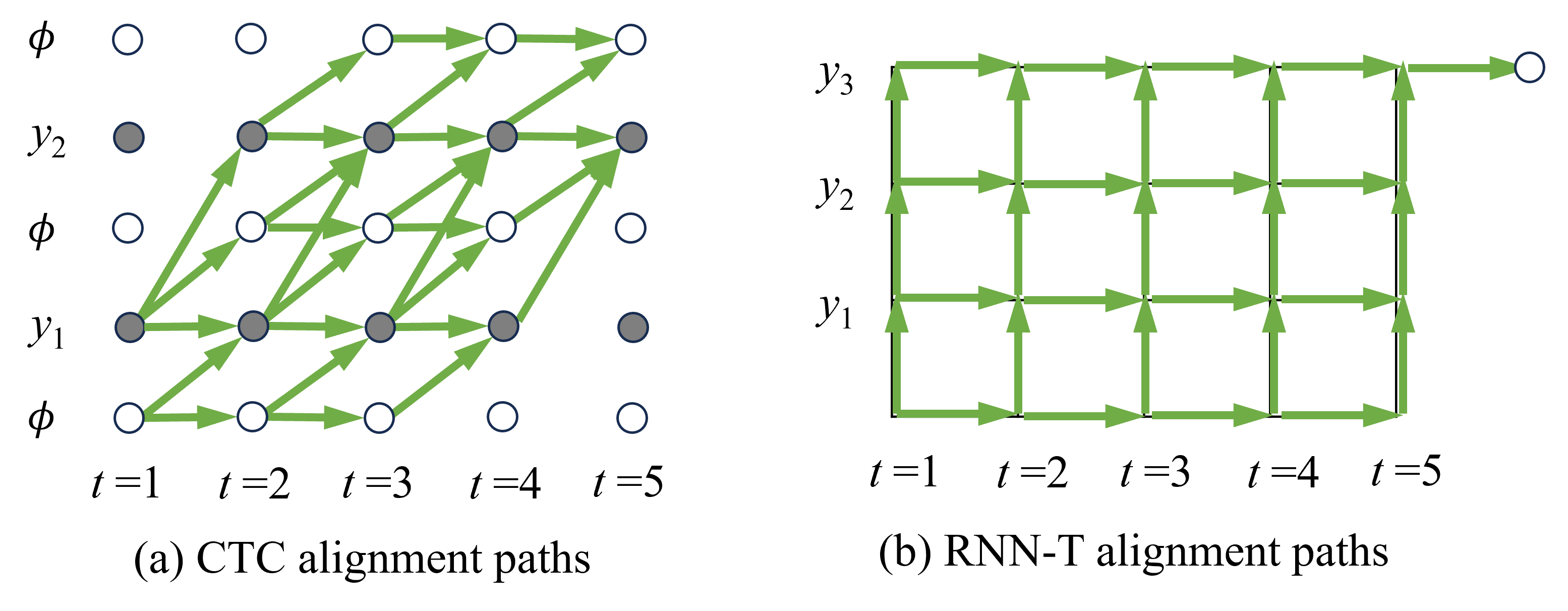} 
        \end{minipage}
        \centering
    \vspace*{-4mm}
    \caption{Alignment paths of CTC and RNN-T. The CTC alignment path is of length $T$, while the RNN-T alignment path is of length $(T + S)$.} 
    \label{fig:alignment}
\end{figure}

The CTC decoder predicts a $T$-length monotonic alignment sequence $Z_{\text{ctc}} = (z^{\text{ctc}}_t \in \mathcal{V} \cup \{\phi\} \mid t = 1, ..., T)$ in a time-synchronous manner, where $\phi$ denotes a blank token. 
Each alignment sequence is deterministically mapped to a corresponding output sequence by removing all blank tokens $\phi$ and repetitive tokens defined by $\mathcal{B_{\text{ctc}}}(Z_{\text{ctc}}) = Y$ (e.g., $\mathcal{B_{\text{ctc}}}([a, a, \phi, a, \phi, b, b]) = [a, a, b]$). 
This mapping strategy constrains the monotonic alignment, reflecting the nature of ASR tasks.
Note that the token sequence $Y$ must be shorter than the $T$-length input sequence $H$.
Instead of directly estimating output posteriors, $P(Y|X)$, the CTC decoder estimates the alignment posteriors, $P(Z_{\text{ctc}}|X)$ as follows: 
\begin{equation}
\label{eq:ctc-alignment}
    P_{\text{ctc}}(Y \mid X) = \sum_{Z_{\text{ctc}} \in \mathcal{B^{\text{-1}}_{\text{ctc}}}(Y)} P_{\text{}}(Z_{\text{ctc}} \mid X),
\end{equation}
where $\mathcal{B^{\text{-1}}_{\text{ctc}}}(Y)$ is a set of all possible alignment sequences of~$Y$ as shown in Figure~\ref{fig:alignment} (a).
$P(Z_{\text{ctc}}|X)$ is computed by using a probabilistic chain rule and conditional independence assumption as follows:
\begin{equation}
\label{eq:ctc}
    P_{\text{}}(Z_{\text{ctc}} \mid X) = \prod_{t=1}^{T} P\left(z^{\text{ctc}}_{t} \mid z^{\text{ctc}}_{1:t-1}, \bm{h}_t \right) \approx \prod_{t=1}^{T} P\left(z^{\text{ctc}}_{t} \mid \bm{h}_t \right),
\end{equation}
where the previous alignment sequence $z^{\text{ctc}}_{1:t-1}$ is discarded based on the conditional independence assumption.
While the conditional independence assumption allows parallel computation, resulting in a fast inference, the CTC decoder often leads to suboptimal performance due to the simplified dependencies between output labels.

Given $H$ generated by the encoder in Eq. \eqref{eq:conformer-encoder}, the CTC decoder, typically consisting of a linear layer, estimates the alignment posterior as follows: 
\begin{equation}
    P\left(z^{\text{ctc}}_{t} \mid \bm{h}_t\right) = \mathrm{Softmax}(\mathrm{Linear}(\bm{h}_t)).
\end{equation}
During training, CTC optimizes the model parameters by minimizing the negative log-likelihood as expressed below:
\begin{equation}
L_{\text{ctc}} = - \log P_{\text{ctc}}(Y \mid X).
\label{eq:loss_ctc}
\end{equation}

\vspace*{-0mm}

\subsection{RNN-T}
\label{sec:rnnt}

Similar to the CTC decoder (Eq.~\eqref{eq:ctc-alignment}), the RNN-T decoder predicts a monotonic alignment sequence $Z_{\text{rnnt}} = (z^{\text{rnnt}}_i \in \mathcal{V} \cup \{\phi\} \mid i = 1, ..., T + S)$ as follows: 
\begin{equation}
    P_{\text{rnnt}}(Y \mid X) = \sum_{Z_{\text{rnnt}} \in \mathcal{B^{\text{-1}}_{\text{rnnt}}}(Y)} P(Z_{\text{rnnt}} \mid X),
\end{equation}
where $\mathcal{B^{\text{-1}}_{\text{rnnt}}}(Y)$ is a set of all possible alignment sequences of~$Y$ (Figure~\ref{fig:alignment} (b)). 
Each alignment sequence $Z_{\text{rnnt}}$ is deterministically mapped to a corresponding output sequence $Y$ by removing all blank tokens defined by $\mathcal{B_{\text{rnnt}}}(Z_{\text{rnnt}}) = Y$. 
As with the CTC, this mapping strategy for RNN-T constrains the efficient monotonic alignment.
However, unlike CTC which has conditional independence assumption in Eq. \eqref{eq:ctc}, the RNN-T decoder estimates $P(Z_{\text{rnnt}} \mid X)$ considering the conditional dependency as follows:
\begin{equation}
\label{rnntlikelihood}
    P(Z_{\text{rnnt}} \mid X) = \prod_{i=1}^{T+S} P\left(z^{\text{rnnt}}_i \mid \bm{h}_t, y_{1:s-1}\right),
\end{equation}
where $t$ and $s$ are determined by the alignment path $z^{\text{rnnt}}_i$ on the RNN-T lattice as shown in Figure ~\ref{fig:alignment} (b).
Unlike CTC, the RNN-T alignment paths have length $(T + S)$, allowing a longer token sequence $Y$ than the $T$-length input sequence $H$.

The RNN-T decoder typically comprises a prediction network and a joint network. 
The prediction network generates a high-level representation $\bm{g}_s \in \mathbb{R}^{d'}$ by conditioning on the previous non-blank token sequence $y_{1:s-1}$ with an embedding layer as follows: 
\begin{equation}
\label{eq:prediction}
    \bm{g}_s = \mathrm{PredNet}(\mathrm{Embedding}(y_{1:s-1})).
\end{equation}
The joint network estimates the alignment posterior $z^{\text{rnnt}}_i$ by combining $\bm{h}_t$ (Eq.~\eqref{eq:conformer-encoder}) and $\bm{g}_{s}$ (Eq.~\eqref{eq:prediction}) as follows:
\begin{equation}
   P(z^{\text{rnnt}}_i \mid \bm{h}_t, y_{1:s-1}) = \mathrm{Softmax}(\mathrm{JointNet}(\bm{h}_t, \bm{g}_{s})).
\end{equation}
RNN-T optimizes the model parameters by minimizing the negative log-likelihood as described below, 
\begin{equation}
L_{\text{rnnt}} = - \log P_{\text{rnnt}}(Y \mid X).
\label{eq:loss_rnnt}
\end{equation}

\subsection{Attention-based encoder-decoder}
\label{sec:attention}

While CTC and RNN-T predict the alignment sequences (Eqs.~\eqref{eq:ctc} and (\ref{rnntlikelihood})), the attention-based encoder-decoder directly estimates the posterior, $P_{\text{att}}(Y \mid X)$, in an autoregressive manner as follows:
\begin{equation}
\label{attlikelihood}
    P_{\text{att}}(Y \mid X) = \prod_{s=1}^{S} P\left(y_{s} \mid y_{1:s-1}, H\right).
\end{equation}
Specifically, the token history $y_{1:s-1}$ is first converted to the token embeddings and fed into the $M$ transformer blocks with the hidden state sequence $H$ in Eq. \eqref{eq:conformer-encoder}. Then, $P \left(y_{s} \mid y_{1:s-1}, X\right)$ is calculated as follows:
\begin{align}
    \label{eq:transformer}
    & \bm{r}_{s} = \mathrm{Transformer}_s(\mathrm{Embedding}(y_{1:s-1}), H),\\
    & P \left(y_{s} \mid y_{1:s-1}, X\right) = \mathrm{Softmax}(\mathrm{Linear}(\bm{r}_{s})),
\end{align}
where $\bm{r}_{s} \in \mathbb{R}^{d'}$ represents the $s$-th hidden state vector.

Each transformer block in Eq.~\eqref{eq:transformer} comprises an MHSA layer, a multiheaded cross-attention layer (MHCA), and a FFN layer with LN layers and residual connections as follows:
\begin{align}
    &V^{\prime}_m = {V_m} + \mathrm{MHSA}(\mathrm{LN}({V_m})),\\
    \label{eq:mhca}
    &V^{\prime\prime}_m = V^{\prime}_m + \mathrm{MHCA}(\mathrm{LN}(V^{\prime}_m), H),\\
    &V^{\prime\prime\prime}_m = V^{\prime\prime}_m + \mathrm{FFN}(\mathrm{LN}(V^{\prime\prime}_m)),
\end{align}
where $V_m$ and $V^{\prime\prime\prime}_m$ represent the input and output of the $m$-th transformer block, respectively. 

The MHCA in Eq.~\eqref{eq:mhca} is stacked cross-attention layers, and each cross-attention layer can be formulated as follows:
\begin{align}
   & \mathrm{CrossAttn}(V^{\prime}_m, H) = \mathrm{Softmax}\left(\frac{W^q V^{\prime}_m (W^k H)^{T}}{\sqrt{d^{\prime}}} \right) W^v H,
\label{eq:audioattn}
\end{align}
where $W^q$, $W^k$, and $W^v$ represent the projection weight matrices for the query, key, and \textcolor{black}{value}, respectively. Here, the softmax term represents the alignment. Unlike CTC and RNN-T, which have reasonable monotonic alignment properties as discussed in Section~\ref{sec:ctc} and \ref{sec:rnnt}, the attention-based encoder-decoder does not maintain this constraint.
While approaches such as monotonic multihead attention have been proposed to enforce monotonic constraints within attention models \cite{ma2019monotonic,inaguma20b_interspeech}, the standard attention mechanism lacks this property and is more prone to alignment errors. Thus, the attention-based encoder-decoder may produce less consistent alignments compared to CTC and RNN-T.

Attention-based encoder-decoder optimizes the model parameters by minimizing the negative log-likelihood described as follows:
\begin{equation}
    L_{\text{att}} = - \log P_{\text{att}}(Y \mid X).
\label{eq:loss_att}
\end{equation}

\subsection{Mask-CTC}
\label{sec:maskctc}

Unlike the attention-based encoder-decoder model (Eq.~\eqref{attlikelihood}), the Mask-CTC estimates the entire token sequence in a non-autoregressive manner based on a masked language model (MLM)~\cite{ghazvininejad2019mask}.
Specifically, the MLM predicts a set of masked tokens $y_{\text{mask}}$ conditioning on the hidden state sequence $H$ in Eq.~\eqref{eq:conformer-encoder} and observed (unmasked) tokens $y_{\text{obs}}$ as follows:
\begin{equation}
    P_{\text{mlm}}(y_{\text{mask}} \mid y_{\text{obs}}, X) = \prod_{y \in y_\text{mask}} P_{\text{}}(y \mid y_{\text{obs}}, H).
\end{equation}
During training, randomly sampled tokens from the reference transcription are masked, while during inference, the low confidence tokens of the greedy CTC results are masked to generate $y_{\text{mask}}$.

Specifically, the observed tokens $y_{\text{obs}}$ are first converted to the token embeddings and fed into the $M$ transformer blocks with hidden state sequence $H$ in Eq.~\eqref{eq:conformer-encoder} as follows:
\begin{align}
    & r_{\text{mask}} = \mathrm{Transformer_{mask}}(\mathrm{Embedding}(y_{\text{obs}}), H),\\
    & P_{\text{mlm}}(y_{\text{mask}} \mid y_{\text{obs}}, X) = \mathrm{Softmax}(\mathrm{Linear}(r_{\text{mask}})),
\end{align}
where $r_{\text{mask}}$ represents the hidden state vectors.
Unlike the attention-based encoder-decoder model, the MLM predicts arbitrary subsets of masked tokens $y_{\text{mask}}$ in a non-autoregressive manner. This mask-based approach maintains fast parallelized inference, but unlike CTC, it accounts for label dependencies. 

Mask-CTC optimizes the model parameters by minimizing the following negative log-likelihood as follows:
\begin{equation}
    L_{\text{mlm}} = - \log P_{\text{mlm}}(y_{\text{mask}}|y_{\text{obs}}, X).
\label{eq:loss_mlm}
\end{equation}

\vspace*{1mm}
\section{Proposed 4D ASR model}
\label{sec:proposed}
\vspace*{1mm}
This section describes the joint training process utilizing the two-stage optimization strategy and the beam search methodology employed in the proposed 4D model.

\subsection{Joint training with a two-stage optimization strategy}
\label{sec:jointraining}

The joint training is performed using the weighted sum of losses in Eqs.~\eqref{eq:loss_ctc}, \eqref{eq:loss_rnnt}, \eqref{eq:loss_att}, and \eqref{eq:loss_mlm} described as follows:
\begin{align}
    &L = \lambda_{\text{ctc}} L_{\text{ctc}} + \lambda_{\text{rnnt}} L_{\text{rnnt}} + \lambda_{\text{att}} L_{\text{att}} + \lambda_{\text{mlm}} L_{\text{mlm}},\\
    \label{eq:multitask}
    &\lambda_{\text{ctc}} + \lambda_{\text{rnnt}} + \lambda_{\text{att}} + \lambda_{\text{mlm}} = 1.0,
\end{align}
where $(\lambda_{\text{ctc}}, \lambda_{\text{rnnt}}, \lambda_{\text{att}}, \lambda_{\text{mlm}})$ represent training weights.
The training weights are usually determined experimentally \cite{watanabe2017hybrid} or based on meta-learning \cite{Lin2019AdaptiveAT}. 
The complexity introduced by four weights makes conducting exhaustive experimental exploration impractical. 
To address this challenge, we implement a two-stage optimization strategy to determine the training weights. In the first stage, all four training weights are initialized to be equal $(0.25, 0.25, 0.25, 0.25)$. Subsequently, in the second stage, these weights are proportionally adjusted based on the epochs at which each validation loss reaches its minimum value in the first stage. 
For example, if the validation losses ($L_{\text{ctc}}, L_{\text{rnnt}}, L_{\text{att}}, L_{\text{mlm}}$) reach their minimum values at the 10th, 10th, 10th, and 70th epochs in the first stage, the training weights in the second stage are set to $(0.1, 0.1, 0.1, 0.7)$. 
Each validation loss is normalized by its value at the end of the first epoch to prevent any single loss from dominating backpropagation due to differences in scale or variance. This normalization ensures balanced dynamic ranges and variances across the four losses. 
In the second stage, training begins with a flat start, rather than continuing from the end of the first stage. These weights remain fixed throughout each stage of training.
This strategy is based on the proposition that losses requiring more epochs to converge should be given higher weights.

\subsection{Overview of the joint decoding of CTC/RNN-T/attention}
\label{sec:joint docoder}

Another contribution of this paper is the introduction of three joint beam search algorithms using CTC/RNN-T/attention decoders: attention-driven, CTC-driven, and RNN-T-driven joint beam search. 
The proposed joint beam search algorithms consist of three parts: hypotheses expansion using a primary decoder, joint scoring using the other two decoders, and beam pruning based on the joint score.
The difference between the three joint beam search algorithms is based on which decoder is the primary decoder.
Subsequent sections delve into the specifics of the three variants of the proposed joint beam search algorithms. 

Note that the Mask-CTC is excluded from the proposed beam search because it predicts the entire output sequence in parallel, unlike the other three decoders. 

\vspace*{-0mm}
\subsection{Attention-driven joint beam search}
\label{sec:attn-driven}

In the attention-driven joint beam search, the attention decoder serves as the primary decoder to generate hypotheses whose scores are augmented by the CTC decoder and the RNN-T decoder as shown in Algorithm 1. This approach directly extends the joint CTC/attention decoding algorithm proposed in \cite{watanabe2017hybrid}, including RNN-T scoring.

\subsubsection{Overall flow}
Similar to the joint CTC/attention decoding, the attention decoder primarily generates $k_\text{pre}$ hypotheses (\textit{ext\_hyps}) with the scores $\alpha_{\text{att}}$ label-synchronously (lines 5-10 in Algorithm 1), where $k_\text{pre}$ represents the prebeam size, which is larger than the main beam size $k_\text{beam}$. 
Subsequently, each hypothesis $\tilde{l}$ in \textit{ext\_hyps} is scored using the CTC and RNN-T decoders (lines 13-14 in Algorithm 1). The CTC score $\alpha_{\text{ctc}}$ is computed using the CTC prefix scoring as in \cite{watanabe2017hybrid}. 
Similarly, we propose the RNN-T prefix scoring to calculate the RNN-T score $\alpha_{\text{rnnt}}$ label-synchronously, which will be described in Section \ref{sec:rnntprefix}. 
Each score is then added using the decoder weights ($\mu_{\text{ctc}}$, $\mu_{\text{att}}$, $\mu_{\text{rnnt}}$) along with the length penalty $\beta$ (line 16 in Algorithm 1).
After that, if $y_s$ = $<$eos$>$, the hypothesis $\tilde{l}$ is assumed to be complete and stored in \textit{end\_hyps} (line 18 in Algorithm 1). 
Otherwise, $\tilde{l}$ is stored in \textit{ext\_hyps} with the joint score $\alpha_{\text{joint}}$ (line 20 in Algorithm 1).
Finally, the top $k_\text{beam}$ hypotheses (\textit{hyps}) are retained for the next label frame based on the joint score $\alpha_{\text{joint}}$ (line 22 in Algorithm 1). 
\textcolor{black}{After completing the beam search process, the best hypothesis is selected from \textit{end\_hyps} (line 25 in Algorithm 1).}

\begin{figure}[t!]
    \centering
        \begin{minipage}{0.49\textwidth}
            \includegraphics[width=\textwidth]{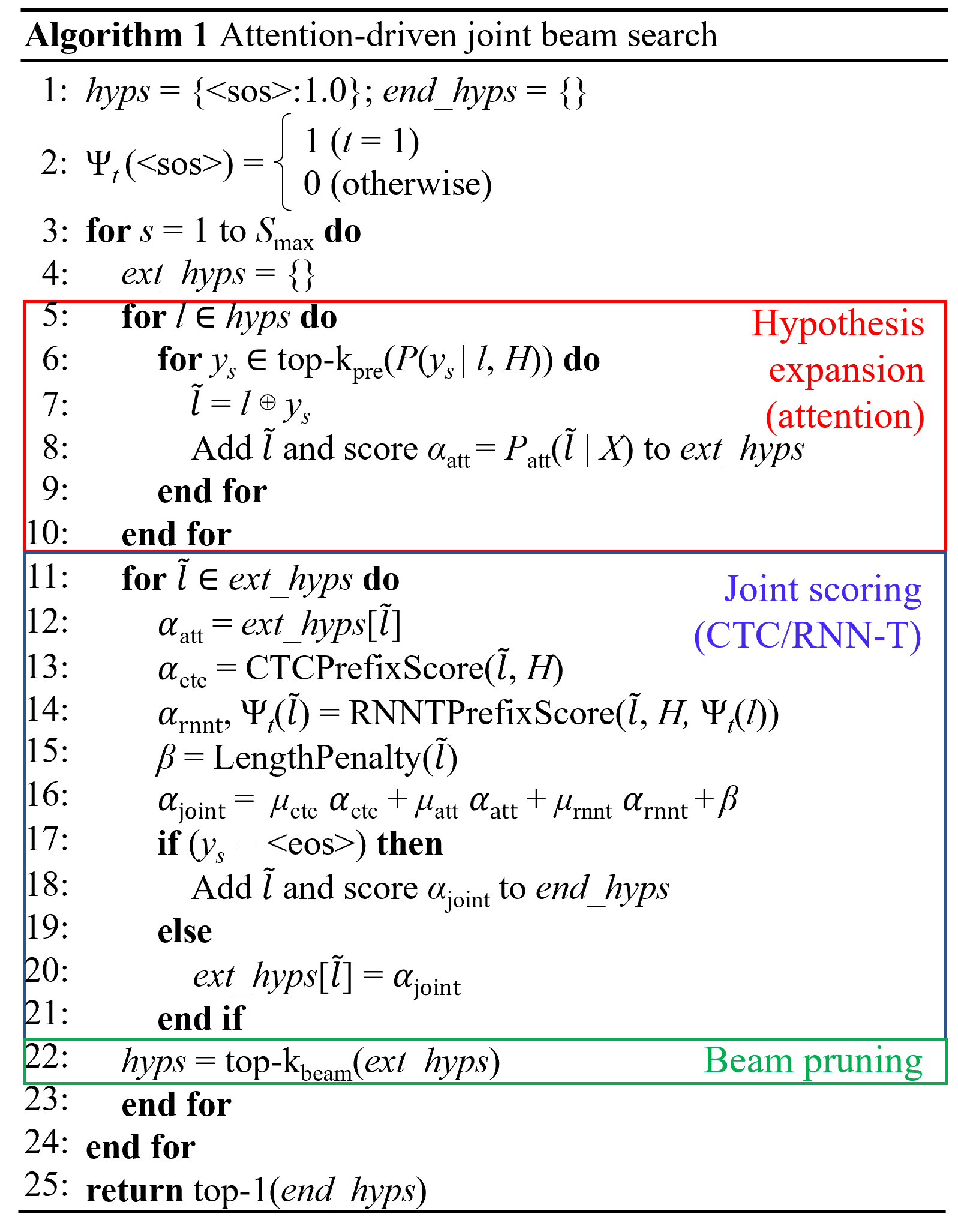} 
        \end{minipage}
        \centering
    \label{fig:algorithm1}
\vspace*{-6mm}
\end{figure}

\subsubsection{RNN-T prefix scoring}
\label{sec:rnntprefix}

\textcolor{black}{
The RNN-T prefix scoring (RNNTPrefixScore($\cdot$) in line 14 of Algorithm 1) computes the RNN-T score $\alpha_{\text{rnnt}}$ from the hypothesis $\tilde{l}$ and the hidden state sequence $H$ (Eq.~\eqref{eq:conformer-encoder}). It introduces an internal variable $\Psi_t(\cdot)$ to cache the previous prefix scores for efficient computation.
}

\begin{figure}[t!]
    \centering
        \begin{minipage}{0.48\textwidth}
            \includegraphics[width=\textwidth]{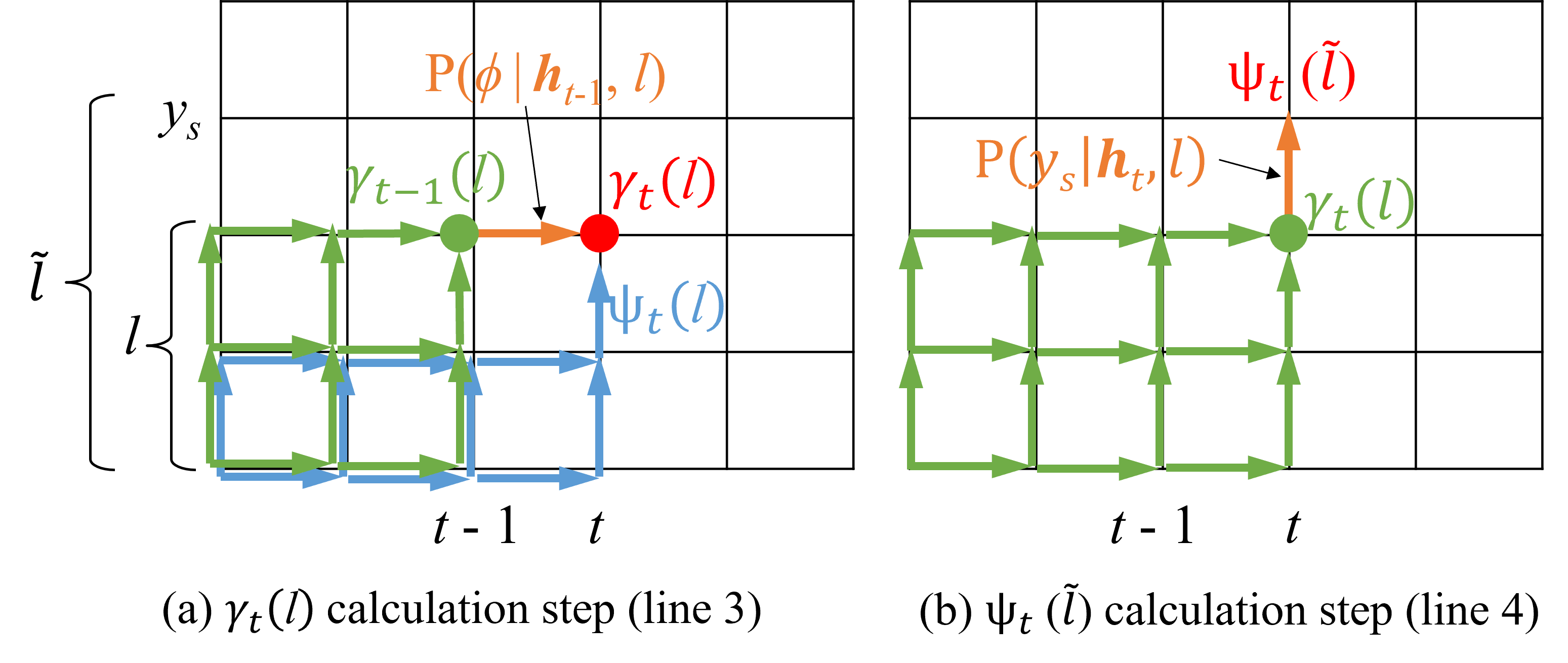} 
        \end{minipage}
        \centering
    \vspace*{-2mm}
    \caption{RNN-T prefix scoring (lines 3 and 4 in Algorithm 2).} 
    \label{fig:rnnt_prefix}
\vspace*{-2mm}
\end{figure}

Algorithm 2 shows the details of the RNN-T prefix scoring, which computes the sum of the probabilities of all RNN-T paths that yield the hypothesis $\tilde{l}$. 
Specifically, it repeats the $\gamma_t(l)$ computation step and the $\Psi_t(\tilde{l})$ computation step (lines 3-4 in Algorithm 2). Here, $\gamma_t(l)$ and $\Psi_t(\tilde{l})$ denote the sum of the probabilities yielding hypothesis $l$ and the probability of outputting the last token $y_s$, resulting in the hypothesis $\tilde{l}$ at time $t$, respectively (Figure \ref{fig:rnnt_prefix}).

First, $\gamma_t(l)$ is computed based on $\Psi_t(l)$, which is initialized for $t = 1, \cdots ,T$ before starting the beam search (line 2 in Algorithm 1). 
As depicted in Figure \ref{fig:rnnt_prefix} (a), $\gamma_t(l)$ is the sum of $\Psi_t(l)$ and the probability of outputting a blank label $\phi$ from $\gamma_{t-1}(l)$. 
Then, $\Psi_t(\tilde{l})$ is computed using $\gamma_t(l)$ (line 4 in Algorithm 2). As shown in Figure \ref{fig:rnnt_prefix} (b), $\Psi_t(\tilde{l})$ is the probability of outputting $y_s$ from $\gamma_t(l)$.
These processes are computed iteratively for $t = 1, \cdots, T$.
Finally, $\Psi_t(\tilde{l})$ is summed for $t = 1, \cdots, T$ (line 9 in Algorithm 2). If $y_s$ is the $<$eos$>$ token, the probability of outputting a blank label $\phi$ at time $T$ is determined from $\gamma_T(l)$ (line 7 in Algorithm 2).

\begin{figure}[t!]
    \centering
        \begin{minipage}{0.485\textwidth}
            \includegraphics[width=\textwidth]{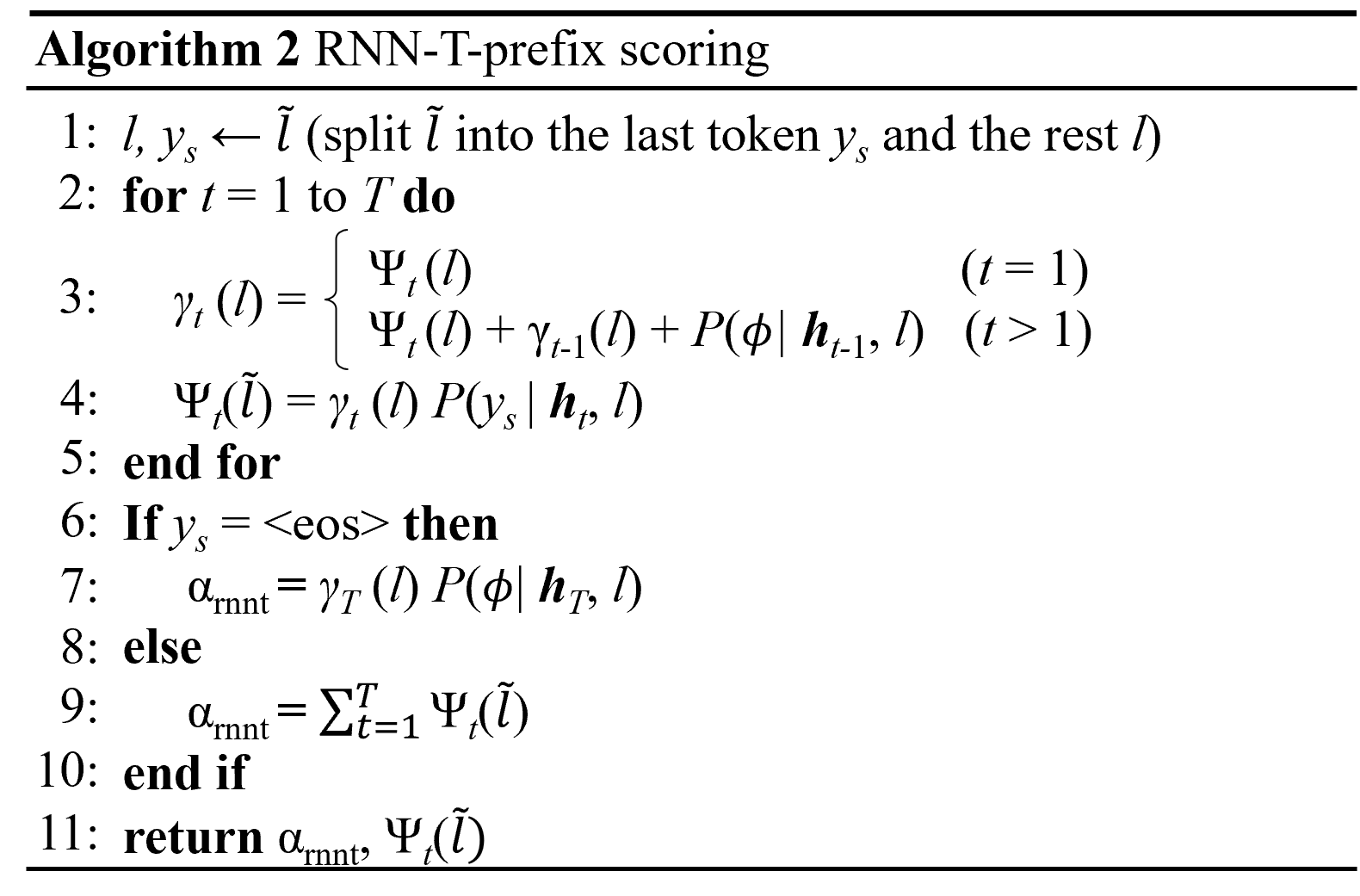} 
        \end{minipage}
        \centering
    \label{fig:algorithm2}
\vspace*{-2mm}
\end{figure}

Note that CTC prefix scoring and RNN-T prefix scoring require more computation for scoring than using the CTC and RNN-T decoders as the primary decoder, respectively, because all possible CTC and RNN-T alignment paths must be computed.

\subsection{CTC-driven joint beam search}
\label{sec:ctc-driven}

The CTC-driven joint beam search employs the CTC decoder as the primary decoder, and the scores of the hypotheses generated by the CTC decoder are augmented by the attention decoder and the RNN-T decoder. This approach is similar to CTC-driven beam search with the attention decoder \cite{moritz2019triggered, yan2022ctc}, but with an additional consideration of RNN-T likelihoods.

As shown in Algorithm 3, the CTC decoder generates the $k_\text{pre}$ hypotheses with the CTC score $\alpha_{\text{ctc}}$, 
\textcolor{black}{
which are stored in \textit{ext\_hyps} after removing blank labels
}
(lines 4-9 in Algorithm 3).
Subsequently, 
\textcolor{black}{
each non-blank hypothesis $\tilde{l}$ in \textit{ext\_hyps} is} 
scored using the attention and RNN-T decoders (lines 12-13 in Algorithm 3). 
The attention score $\alpha_{\text{att}}$ is computed using the forward computation as described in Eq.~(\ref{attlikelihood}). Similar to the attention-driven joint beam search (in Algorithm 1), the RNN-T prefix scoring is used to compute the RNN-T score $\alpha_{\text{rnnt}}$. 
\textcolor{black}{
RNNTPrefixScore($\cdot$) takes the non-blank hypothesis $\tilde{l}$ as input and computes the probabilities of all possible alignment paths over $t = 1, \cdots, T$ (Figure~\ref{fig:rnnt_prefix}). 
This approach addresses the difference in optimal alignment paths between CTC and RNN-T (Figure~\ref{fig:alignment}), which prevents direct scoring of CTC-generated alignment paths with the RNN-T decoder.
The third argument, $\Phi_t(l)$, is an internal variable used to cache the previous prefix scores for $\tilde{l}$.
}
Each decoder score is then added using the decoder weights ($\mu_{\text{ctc}}$, $\mu_{\text{att}}$, $\mu_{\text{rnnt}}$) along with the length penalty (line 15 in Algorithm 3).
Finally, the top $k_\text{beam}$ hypotheses, \textit{hyps}, are retained for the next time frame based on the joint score (line 17 in Algorithm 3). Unlike the attention-driven joint beam search, this process is iterated time-synchronously ($t=1, \cdots, T$).

\begin{figure}[t!]
    \centering
        \begin{minipage}{0.49\textwidth}
            \includegraphics[width=\textwidth]{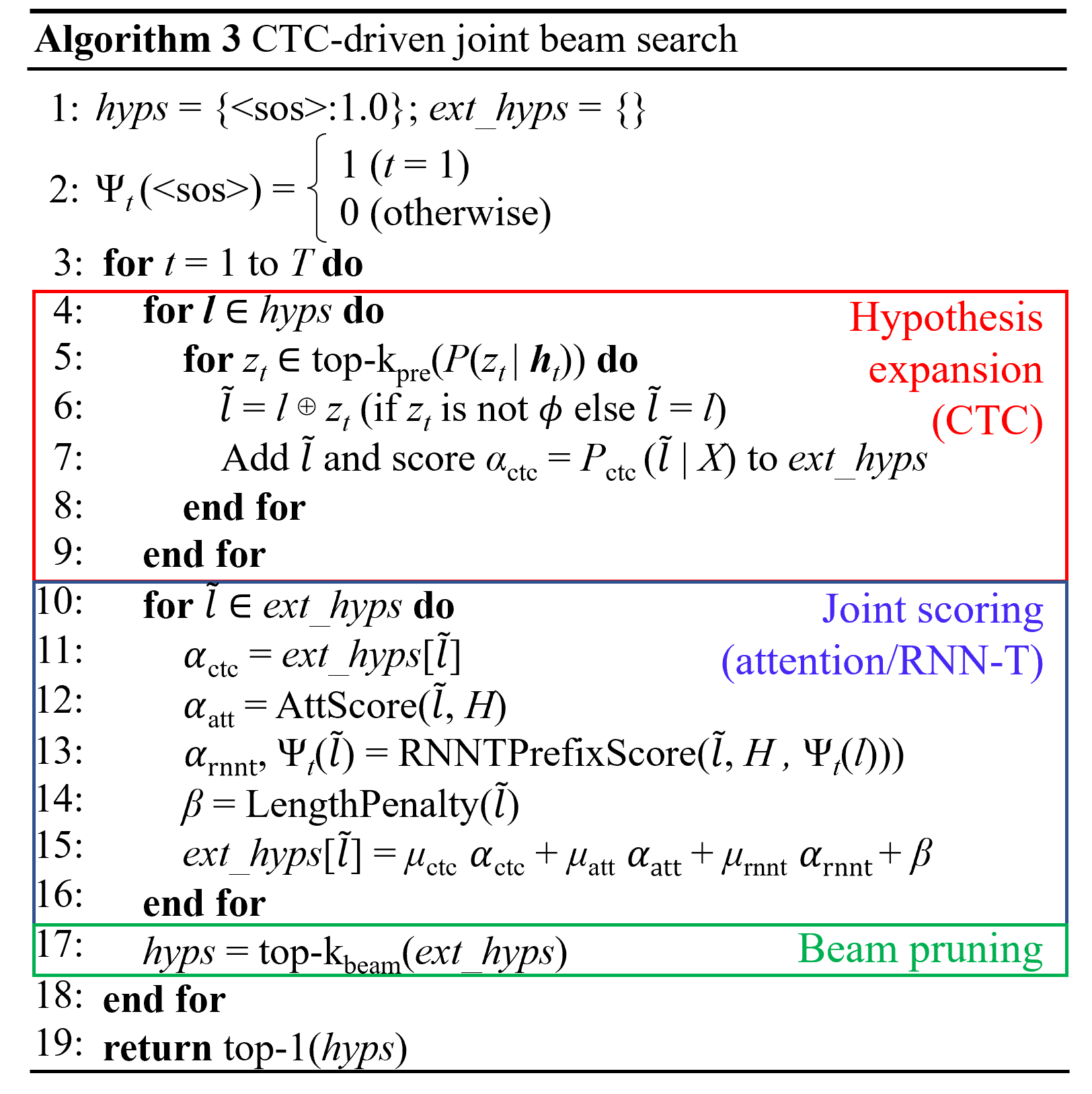} 
        \end{minipage}
        \centering
    \label{fig:algorithm3}
\vspace*{-5mm}
\end{figure}

Note that while the CTC-driven joint beam search algorithm avoids CTC prefix scoring which calculates all possible CTC alignment paths, it still computes all possible RNNT alignment paths via RNNTPrefixScore($\cdot$),
\textcolor{black}{
which increases the computational cost.
}

\subsection{RNN-T-driven joint beam search}
\label{sec:rnnt-driven}

The RNN-T-driven joint beam search employs the RNN-T decoder as the primary decoder, and the scores of the hypotheses generated by the RNN-T decoder are augmented by the CTC decoder and attention decoder. Unlike the two-pass rescoring-based methods \cite{sainath2019two,hu2021transformer,tian2022hybrid,wang2022deliberation,yao2021wenet}, the joint scoring is performed time-synchronously in the joint beam search.

First, the RNN-T decoder generates the $k_\text{pre}$ hypotheses (lines 4-9 in Algorithm 4).
Subsequently, the generated hypotheses, \textit{ext\_hyps}, are scored by combining the CTC and attention decoders (lines 12-13 in Algorithm 4).
The attention score is calculated using the forward computation as in Eq.~(\ref{attlikelihood}).
The CTC score is calculated using  CTC prefix scoring as in \cite{watanabe2017hybrid}.
Note that although both CTC and RNN-T are based on time-synchronous beam search, they have different alignment paths and different mapping functions, as shown in Figure~\ref{fig:alignment}. Since the RNN-T alignment path cannot be directly scored by the CTC decoder, the CTC prefix scoring is used after all blank labels are removed.
Each decoder score is then added using the decoder weights ($\mu_{\text{ctc}}$, $\mu_{\text{att}}$, $\mu_{\text{rnnt}}$) along with the length penalty (line 15 in Algorithm 4). 
Finally, the top $k_\text{beam}$ hypotheses, \textit{hyps}, are retained for the next time frame based on the obtained joint score (line 17 in Algorithm 4). Similar to the CTC-driven joint beam search, this process is repeated time-synchronously ($t=1, \cdots, T$).

\begin{figure}[t!]
    \centering
        \begin{minipage}{0.49\textwidth}
            \includegraphics[width=\textwidth]{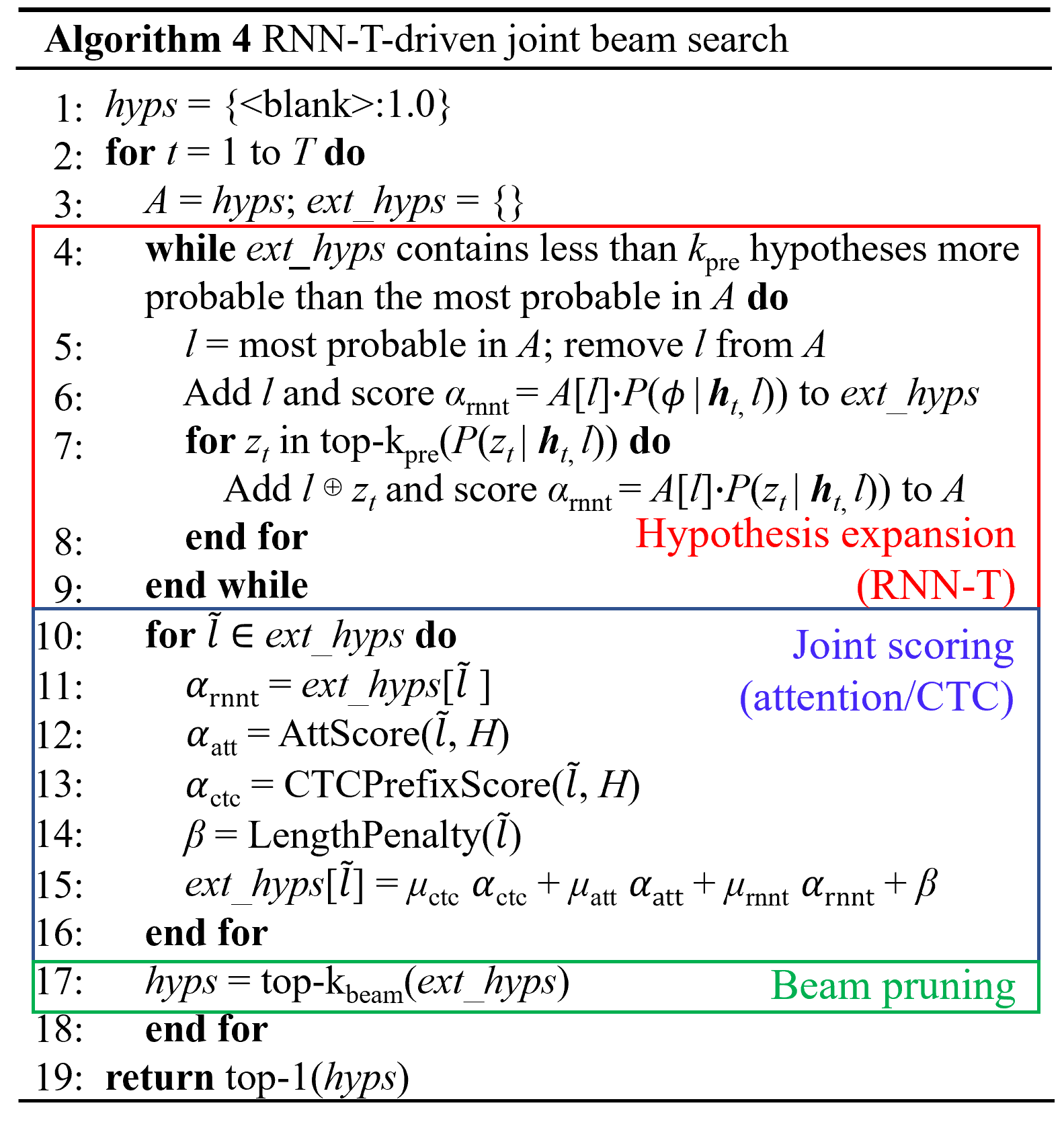} 
        \end{minipage}
        \centering
    \label{fig:algorithm4}
\vspace*{-3mm}
\end{figure}

Note that the RNN-T-driven joint beam search algorithm avoids RNN-T prefix scoring which computes all RNN-T alignment paths unlike the other two joint beam search algorithms (Section \ref{sec:attn-driven} and \ref{sec:ctc-driven}).

\section{Experimental setup}
\label{sec:experimental condition}

To evaluate the effectiveness of the proposed method, the 4D model is trained and evaluated using several datasets described below.

\vspace*{-0mm}
\subsection{Model}

The input features are 80-dimensional Mel-scale filter-bank features with a window size of 512 samples and a hop length of 160 samples. The sampling frequency is 16 kHz. Then, SpecAugment \cite{specaug} is applied. 
The conformer encoder comprises two convolutional layers and 12 conformer blocks, each incorporating layer normalization with residual connections. 
The CTC decoder is a linear layer. The attention and MLM decoders have six transformer blocks, each with 2048 linear units. The RNN-T decoder employs a long short-term memory (LSTM) and a linear layer of 640 joint sizes for the prediction and joint networks, respectively.
The details of the model architectures are summarized in Table \ref{condition}.

\begin{table}
\caption{Model configurations \textcolor{black}{during training, including the parameter counts for each decoder. Unused decoder parameters are ignored during decoding.}
}
\vspace{-5mm}
\label{condition}
\begin{center}
\begin{tabular}{@{}cc|ccc}
\hline
   &  & \multicolumn{2}{c}{LibriSpeech} & In-house\\
   &  & 960 hours & 100 hours & 855 hours\\
\hline
  & \textbf{Convolution} \\
  & Kernel size &  \multicolumn{3}{c}{(3 x 3), (5 x 5)}\\
 & Stride & (2, 2) & (2, 2) & (2, 3)\\
 & \textbf{Conformer}  \\
\textbf{Encoder}  & Layers & 12 & 12 & 12 \\
 & Heads & 8 & 4 & 8 \\
 & Dimension & 512 & 256 & 512 \\
 & Linear units & 2048 & 1024 & 2048 \\
 & \textcolor{black}{\textbf{Params M}} & \textcolor{black}{83.3} & \textcolor{black}{20.9} & \textcolor{black}{83.9} \\
\hline
 & \textbf{Linear}\\
\textbf{CTC} & Layers & 1 & 1 & 1\\
 & \textcolor{black}{\textbf{Params M}} & \textcolor{black}{2.6} & \textcolor{black}{1.3} & \textcolor{black}{1.8}\\
\hline
\multirow{6}{*}{\textbf{Attn}} & \textbf{Transformer} \\
  & Layers & 6 & 6 & 6\\
  & Heads & 8 & 4 & 8 \\
  & Dimension & 512 & 256 & 512 \\
  & Linear units & 2048 & 2048 & 2048 \\
  & \textcolor{black}{\textbf{Params M}} & \textcolor{black}{30.3} & \textcolor{black}{12.0} & \textcolor{black}{29.0} \\
\hline
\multirow{6}{*}{\textbf{Mask-CTC}} & \textbf{Transformer} \\
  & Layers & 6 & 6 & 6 \\
  & Heads & 8 & 4 & 8 \\
  & Dimension & 512 & 256 & 512 \\
  & Linear units & 2048 & 2048 & 2048 \\
  & \textcolor{black}{\textbf{Params M}} & \textcolor{black}{30.3} & \textcolor{black}{12.0} & \textcolor{black}{29.0} \\
\hline
\multirow{6}{*}{\textbf{RNN-T}} &  \textbf{LSTM} \\
  & Layers & 1 & 1 & 1 \\
  & Hidden size & 512 & 256 & 512 \\
  & \textbf{Linear}\\
  & Joint size & 640 & 640 & 640 \\
  & \textcolor{black}{\textbf{Params M}} & \textcolor{black}{8.5} & \textcolor{black}{5.4} & \textcolor{black}{8.5} \\
\hline
\multicolumn{2}{c|}{\textbf{\textcolor{black}{Total} params} M} & 155.0 & 51.6 & 152.2 \\
\hline
\end{tabular}
\end{center}
\vspace*{-3mm}
\end{table}

The proposed model is trained for 150 epochs using the Adam optimizer \cite{kingma2014adam} at a learning rate of 0.0015, with 15000 warmup steps. 
Four GPUs are used during training, and the batch size is dynamically adjusted based on the input length using the numel batch type in the ESPnet toolkit \cite{espnet}. In our experiments, the average batch size was 47.
The training weights ($\lambda_{\text{ctc}}$, $\lambda_{\text{rnnt}}$, $\lambda_{\text{att}}$, $\lambda_{\text{mlm}}$) of the second stage are (0.15, 0.10, 0.30, 0.45) based on the two-stage optimization strategy described in Section \ref{sec:jointraining}, which will be discussed in Section \ref{sec:joint training}. The decoder weights ($\mu_{\text{ctc}}$, $\mu_{\text{rnnt}}$, $\mu_{\text{att}}$) of the CTC-driven, RNN-T-driven, and attention-driven beam search in Algorithms 1, 3, and 4 are (0.3, 0.3, 0.4), (0.1, 0.4, 0.5), and (0.1, 0.4, 0.5), respectively. The main beam size $\text{k}_{\text{beam}}$ and prebeam size $\text{k}_{\text{pre}}$ are 20 and 30, respectively.

In addition, we evaluate the impact of external LM shallow fusion on the proposed 4D model. The LM consists of 16 transformer blocks, each containing 2048 linear units and eight multi-head attention mechanisms with a dimension of 512. The LM is trained for 50 epochs using the Adam optimizer at a learning rate of 0.0025, with 25000 warmup steps.

\subsection{Dataset}

The proposed method is tested using the LibriSpeech (960 hours, 100 hours) \cite{panayotov2015librispeech} and our in-house dataset.
Our in-house dataset\footnote{Our in-house dataset is not released for privacy and confidentiality reasons.} comprises 93 hours of Japanese speech data, collected from various scenarios such as meetings and morning assemblies, in addition to 581 hours of the Corpus of Spontaneous Japanese \cite{csj} and 181 hours of Japanese speech database developed by the Advanced Telecommunications Research Institute International (ATR-APP)~\cite{KUREMATSU1990357}.
The external LM is trained using an additional 850M words of text from the LibriSpeech dataset.
The word/character error rates~(WER/CER) are calculated for the LibriSpeech and our in-house dataset using the ESPnet toolkit \cite{espnet}.

\section{Results}
\label{sec:experiments}

In this section, we first discuss the main results of the proposed method in Section \ref{sec:mainresults}. Subsequently, we explore the effectiveness of both joint training and decoding in Section \ref{sec:joint training} and \ref{sec:beam search}. Finally, we thoroughly compare the three proposed joint beam search algorithms in Section \ref{sec:decodeanalisys}.

\begin{table*}[t]
\caption{Comparison of the 4D models (B1-10) against respective baselines (A1-7). The best WER/CER ($\downarrow$) result in each comparison is \textbf{bolded}, and the best results overall are further \underline{\textbf{underlined}}. The average absolute improvements ($\Delta$) are also shown.}
\vspace*{-11mm}
\label{maintable}
\begin{center}

\include{main_table}

\end{center}
\vspace*{-4mm}
\end{table*}

\subsection{Main results}
\label{sec:mainresults}

\subsubsection{Main results}
Table \ref{maintable} presents a comprehensive overview of the results, evaluating the effect of both joint training and decoding strategies. 
It also includes the number of parameters used during both decoding and training (denoted as decode/train) for the LibriSpeech 960 hours dataset.
\textcolor{black}{In the 4D model, all four decoders are used during training, whereas during decoding, the parameters of unused decoders (Table \ref{condition}) can be ignored.}

First, we observe that the individual decoder branches of the 4D model exhibit superior performance compared to their non-4D trained counterparts (A1-4 vs. B1-4). Specifically, the 4D model enhances the CTC, attention, RNN-T, and Mask-CTC models, resulting in an overall improvement of 0.3-0.7 absolute WER/CER points.
This improvement also holds for existing integration models, including two-pass rescoring and joint beam search of the CTC/attention models \cite{yao2021wenet,watanabe2017hybrid}, as demonstrated by the comparison between A5-6 and B5-6. The results confirm that joint training significantly contributes to the enhanced performance of the individual decoders. 

Beyond joint training, Table \ref{maintable} also shows the results of the three proposed joint beam search algorithms. These methods consistently outperform the CTC/attention baseline, as indicated by the improvements seen in A6 compared to B7-9. In particular, the RNN-T-driven CTC/RNN-T/attention decoding shows a better performance improvement (an average of 0.5 absolute WER/CER point improvement) than the other two joint beam search methods.
Notably, these improvements are more pronounced on the LibriSpeech 100 hours dataset, suggesting that the 4D approach provides a regularization effect that is particularly beneficial when training data are limited.

\subsubsection{Error analysis}
Table \ref{csid} shows the detailed error analysis on the LibriSpeech 100 test-other set in terms of substitution, deletion, and insertion errors. Joint training improves substitution, deletion, and insertion errors in most cases (A1-4 vs. B1-4). In addition, joint decoding further improves performance, particularly in reducing substitution errors (A5,6 vs. B5-9). Notably, the proposed RNN-T-driven joint decoding method reduces substitution errors by 1.3 points compared to the conventional CTC/attention joint decoding method, resulting in the best WER performance (A6 vs. B9).

\begin{table}[t]
\caption{Error analysis on LibriSpeech 100 test-other. The best WER ($\downarrow$) result in each comparison is \textbf{bolded}, and the best results overall are further \underline{\textbf{underlined}}.}
\vspace*{-8mm}
\label{csid}
\begin{center}
\resizebox {\linewidth} {!} {
\begin{tabular}{@{}clc|ccc|c}
\hline
ID & Model & Primary & Sub & Del & Ins & WER \\
\hline
A1 & Attention & - & 13.6 & 3.7 & 2.1 & 19.4 \\
B1 & 4D (Attn) & -& \textbf{13.1} & \textbf{3.1} & \textbf{1.8} & \textbf{17.9} \\
\hline
A2 & CTC & - & 16.2 & 2.4 & 2.0 & 20.7 \\
B2 & 4D (CTC) & - & \textbf{14.9} & \textbf{2.1} & \textbf{1.9} & \textbf{19.0} \\
\hline
A3 & Mask-CTC & - & 16.3 & 2.3 & 2.2 & 20.8 \\
B3 & 4D (Mask-CTC) & - & \textbf{15.0} & \textbf{2.0} & \textbf{1.9} & \textbf{19.0} \\
\hline
A4 & RNN-T & - & 14.2 & \textbf{2.3} & 1.8 & \textcolor{black}{18.3} \\
B4 & 4D (RNN-T) & - & \textbf{13.4} & 2.6 & \underline{\textbf{1.6}} & \textbf{17.6} \\
\hline
A5 & CTC/Attn (2-pass) \cite{yao2021wenet} & CTC & 13.8 & 2.5 & 1.8 & 18.2 \\
B5 & 4D (CTC/Attn) & CTC & \textbf{13.2} & \textbf{2.0} & 1.8 & \textbf{17.0} \\
\hline
A6 & CTC/Attn \cite{watanabe2017hybrid} & Attn & 14.1 & \textbf{1.7} & 2.0 & 17.8 \\
B6 & 4D (CTC/Attn) & Attn & \textbf{13.2} & 2.0 & \textbf{1.8} & \textbf{17.0} \\
\hline
A6 & CTC/Attn \cite{watanabe2017hybrid} & Attn & 14.1 & 1.7 & 2.0 & 17.8 \\
B7 & 4D (CTC/RNN-T/Attn) & Attn & \textbf{13.2} & \underline{\textbf{1.6}} & \textbf{1.8} & 16.6 \\
B8 & 4D (CTC/RNN-T/Attn) & CTC & \textbf{13.0} & 2.0 & \textbf{1.8} & 16.8 \\
B9 & 4D (CTC/RNN-T/Attn) & RNN-T & \underline{\textbf{12.8}} & 1.9 & \textbf{1.8} & \underline{\textbf{16.4}} \\
\hline
\end{tabular}
}
\end{center}
\vspace*{-0mm}
\end{table}

\subsubsection{Impact of LM shallow fusion}
We also evaluate the impact of LM shallow fusion on the proposed 4D model. Table \ref{lm_fusion} compares the performance of the CTC/attention baseline and the 4D model, both with and without LM fusion, on the LibriSpeech 960 hours dataset. For a fair comparison, the attention-driven joint beam search algorithm is used for both the CTC/attention baseline and the proposed 4D model.

\textcolor{black}{The results show that LM fusion yields a larger performance improvement compared with the 4D joint modeling. Nevertheless, the 4D model with LM fusion achieves WERs comparable to the CTC/attention model with LM fusion. This demonstrates that the 4D model can incorporate LM fusion without degrading performance, while maintaining its flexibility to switch between decoders during inference.}

\begin{table}[t]
\caption{Impact of LM shallow fusion on the LibriSpeech 960 test set.}
\vspace*{-5mm}
\label{lm_fusion}
\begin{center}
\begin{tabular}{@{}lc|cc}
\hline
 &  & \multicolumn{2}{c}{WER ($\downarrow$)} \\
Decoding method & LM & test-clean & test-other \\
\hline
Attn-driven CTC/Attn \cite{watanabe2017hybrid} & - & 2.5 & \textbf{5.2} \\
4D Attn-driven CTC/Attn & - & \textbf{2.4} & 5.3 \\
\hline
Attn-driven CTC/Attn \cite{watanabe2017hybrid} & yes & \textbf{2.0} & 4.4 \\
4D Attn-driven CTC/RNN-T/Attn & yes & \textbf{2.0} & \textbf{4.3} \\
\hline
\end{tabular}
\end{center}
\vspace*{-0mm}
\end{table}

\subsection{Detailed analysis of the joint training}
\label{sec:joint training}

This section provides a detailed analysis of joint training. The subsequent subsections examine the proposed two-stage training, the effect of the number of decoders in joint training, and the training time.

\subsubsection{Analysis of the two-stage training strategy}
\label{sec:two-stage}

\begin{figure}
    \centering
        \begin{minipage}{0.35\textwidth}
        \centering
            \includegraphics[width=1.0\textwidth]{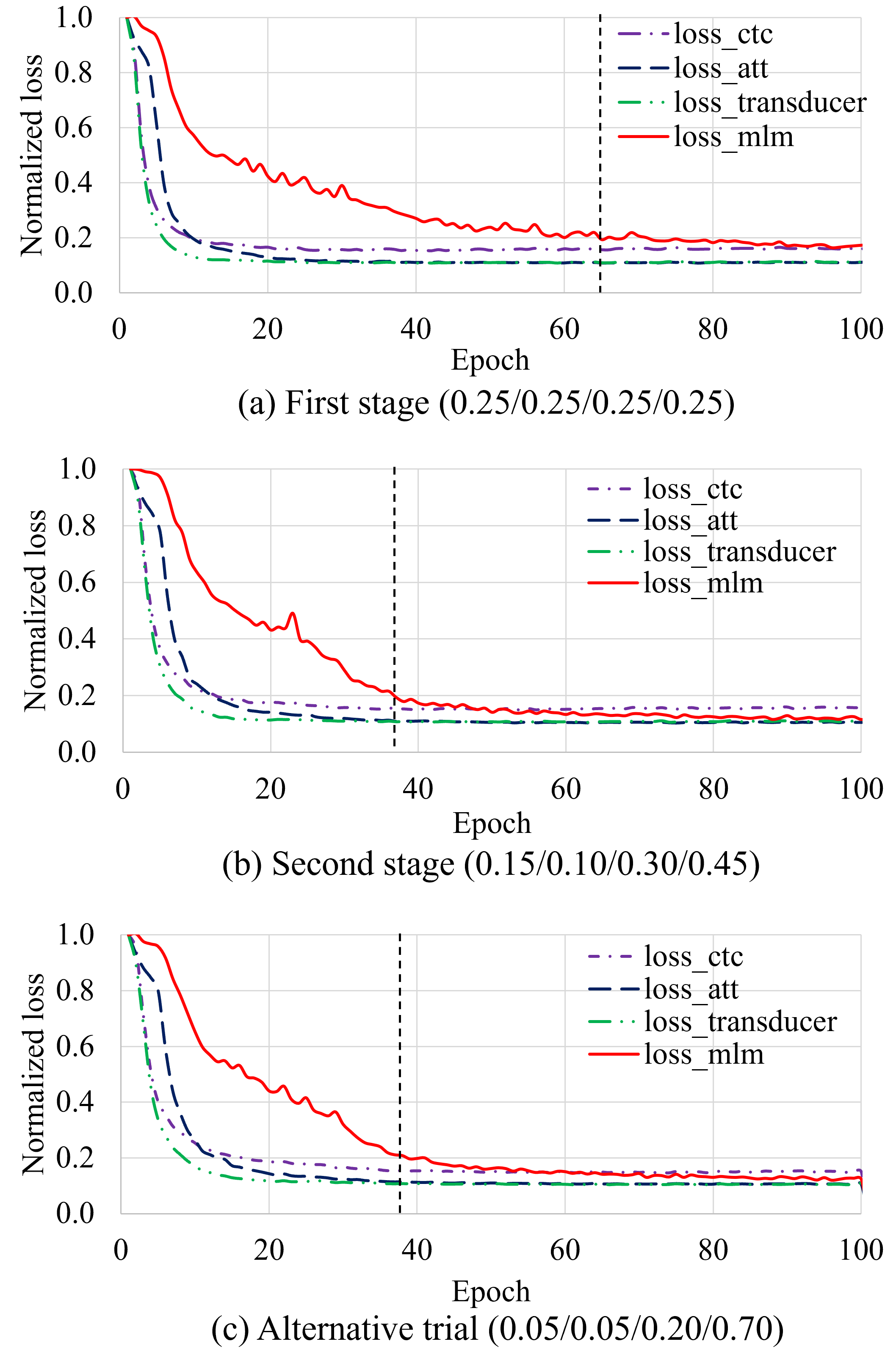} 
        \end{minipage}
        \centering
    \vspace*{-2mm}
    \caption{Normalized validation curves of the first, second training stages, and alternative trial ($\lambda_{\text{ctc}}$, $\lambda_{\text{rnnt}}$, $\lambda_{\text{att}}$, $\lambda_{\text{mlm}}$)}.
    \label{fig:stage12}
\vspace*{-1mm}
\end{figure}

Figure \ref{fig:stage12} illustrates the normalized validation losses in the first and second training stages. Note that the validation losses are normalized for a comparative analysis of convergence speed rather than focusing on the magnitude of the losses. 
During the first training stage (Figure \ref{fig:stage12} (a)), the MLM loss exhibits relatively slower convergence, suggesting potential under-convergence with the MLM decoder or potential overfitting with the other decoders. However, in the second training stage (Figure \ref{fig:stage12} (b)), the gap in convergence speed among the four losses decreases, indicating a more balanced convergence across all decoders.

Table \ref{weights} presents the performance of each decoder on the test-other set of LibriSpeech dataset without and with 4D joint training in the first and second stages.  
Even in the first stage, where each model outperforms the performance of the non-4D joint training model, all four decoders exhibit improved performance in the second stage. 

Despite the improved convergence of the MLM loss in the second stage, it still requires more epochs to converge compared to the other losses. To further explore the effect of increasing the MLM training weight, we conduct an alternative experiment using the training weights ($\lambda_{\text{ctc}}$, $\lambda_{\text{rnnt}}$, $\lambda_{\text{att}}$, $\lambda_{\text{mlm}}$) = (0.05/0.05/0.2/0.7). However, as shown in Figure \ref{fig:stage12} (c), the convergence speed does not improve further, resulting in a degradation in WER as shown in Table \ref{weights}. 
This result may be attributed to the inherent randomness in MLM training, where tokens are randomly masked during training, potentially limiting the convergence speed improvements.

Consequently, while the proposed two-stage approach may not always yield optimal weights, it remains a practical training strategy that can determine effective weights with only two experimental trials.

\begin{table}
\caption{Effect of the proposed two-stage strategy on LibriSpeech (960/100) test-other. 
Non-4D, 1st stage, 2nd stage, and Alternative refer to models trained independently, each stage of the strategy, and the alternative experiment, respectively.}
\vspace{-7.5mm}
\label{weights}
\begin{center}
\resizebox {\linewidth} {!} {
\begin{tabular}{@{}cc|cccc}
\hline
& &  \multicolumn{4}{c}{WER ($\downarrow$) (training weight)} \\
Dataset & Decoder & Non-4D & 1st stage & 2nd stage & Alternative\\
\hline
 & CTC & 6.9 (-) & 6.6 (0.25) & \textbf{6.4} (0.15) & 6.5 (0.05)\\
& RNN-T      & 5.8 (-) & 5.8 (0.25) & 5.7 (0.10) & \textbf{5.6} (0.05)\\
960 hours & Attn  & \textbf{5.6} (-) & 5.7 (0.25) & \textbf{5.6} (0.30) & 6.5 (0.20)\\
 & Mask-CTC   & 6.9 (-) & 7.0 (0.25) & 6.8 (0.45) & \textbf{6.7} (0.70)\\
& Avg & 6.3 & 6.3 & \textbf{6.1} & 6.3\\
\hline
& CTC        & 20.7 (-) & 19.5 (0.25) & \textbf{19.0} (0.15) & 19.4 (0.05)\\
& RNN-T      & 18.3 (-) & 18.1 (0.25) & \textbf{17.6} (0.10) & 17.8 (0.05)\\
100 hours & Attn  & 19.4 (-) & 18.7 (0.25) & \textbf{17.9} (0.30) & 19.3 (0.20) \\
 & Mask-CTC   & 20.8 (-) & 20.4 (0.25) & \textbf{19.0} (0.45) & 19.3 (0.70) \\
& Avg & 19.8 & 19.2 & \textbf{18.4} & 19.0\\
\hline
\end{tabular}
}
\end{center}
\vspace*{-3mm}
\end{table}

\subsubsection{Effect of the number of decoders}
\label{sec:3d4d}

Table \ref{3d4d} depicts the relationship between the number of decoders employed for joint training and the resultant performance. The evaluation aims to investigate the effect of joint training by comparing the proposed 4D model with the CTC/RNN-T/attention (a three-decoder joint model), CTC/attention, and RNN-T models.
To isolate the influence of joint training, the individual decoders are employed for decoding (i.e., CTC, attention, and RNN-T), avoiding the application of joint decoding. As depicted in Table \ref{3d4d}, WER significantly improves as the number of decoders used for joint training increases. This underscores the effectiveness of joint training using multiple decoders with distinct characteristics in effectively regularizing the encoder training.

\begin{table}[t]
\caption{Effect of the number of decoders during \textbf{training} on LibriSpeech (960/100) test-other set.}
\vspace*{-8mm}
\label{3d4d}
\begin{center}
\resizebox {\linewidth} {!} {
\begin{tabular}{@{}lc|ccc}
\hline
 &  & \multicolumn{3}{c}{WER ($\downarrow$)}\\
Joint training   & Decoder & 960 hours & 100 hours & Avg \\
\hline
CTC/Attn &      & 6.9 & 20.7 & 13.8\\
CTC/RNN-T/Attn     & CTC  & 6.8 & 20.6 & 13.7\\
CTC/RNN-T/Attn/Mask-CTC &      & \textbf{6.4} & \textbf{19.0} & \textbf{12.7}\\
\hline
CTC/Attn &  & \textbf{5.6} & 19.4 & 12.5\\
CTC/RNN-T/Attn      & Attention & 6.1 & 18.7 & 12.4\\
CTC/RNN-T/Attn/Mask-CTC  &  & \textbf{5.6} & \textbf{17.9} & \textbf{11.8}\\
\hline
RNN-T  & & 5.8 & 18.3 & 12.0\\
CTC/RNN-T/Attn & RNN-T  & \textbf{5.6} & 18.3 & 12.0\\
CTC/RNN-T/Attn/Mask-CTC &  & 5.7 & \textbf{17.6} & \textbf{11.2}\\
\hline
\end{tabular}
}
\end{center}
\vspace*{-3mm}
\end{table}

\subsubsection{Limitation of joint training}
\label{sec:traintime}

Table \ref{traintime} provides the training time for each model when training single and 4D models on the LibriSpeech 100 hours dataset using NVIDIA A100 80GB GPUs. RNN-T requires substantially more training time than the other two individual models (CTC/Attention and Mask-CTC) due to the considerable amount of GPU memory required for computing all possible RNN-T paths \cite{kuang22_interspeech}. 
Because the 4D model incorporates four decoders, including the RNN-T decoder, the training time naturally increases (roughly the sum of the training times for each model). 

However, during decoding, as long as a single decoder is utilized, there is no need for the forward computation of other decoders, ensuring that the decoding time remains comparable to that of the individual models (as further discussed in Section \ref{sec:rtf-wer}). 
In addition, whereas multiple models are typically trained separately for different application scenarios, the 4D joint model provides a unified framework that can be easily adapted to different application scenarios by simply switching between decoders once 4D joint training is complete.
Thus, although the initial training overhead for the 4D model is higher, it eliminates the need to train models separately for each use case. 
If three individual systems (CTC/attention, RNN-T, and Mask-CTC) need to be trained, the total training time would be 1.21 GPU hours / epoch, meaning that the increase in training time for the 4D model is only 16\%.

\begin{table}
\caption{Training time comparison between individual models and the 4D model. 
As each model uses a different number of GPUs, the comparison is based on GPU hours per epoch.}
\vspace*{-5mm}
\label{traintime}
\begin{center}
\begin{tabular}{@{}c|c|cc}
\hline
Model   & GPUs  & GPU hours / epoch & Increase ($\downarrow$)\\
\hline
CTC/Attn     & 1 & 0.16  & 1.0x\\
Mask-CTC     & 1 & 0.19 & 1.2x\\
RNN-T        & 2 & 0.86 & 5.5x\\
4D           & 4 & 1.40 & 8.9x\\
\hline
\end{tabular}
\end{center}
\vspace*{-3mm}
\end{table}

\subsection{Detailed analysis of the proposed joint beam search}
\label{sec:beam search}

This section provides a detailed analysis of joint decoding, specifically focusing on the RNN-T-driven joint beam search, demonstrating superior performance in Table \ref{maintable}.

\subsubsection{Computational complexity of the beam search}
\label{sec:rtf-wer}

\begin{figure}[t!]
    \centering
        \begin{minipage}{0.49\textwidth}
            \includegraphics[width=\textwidth]{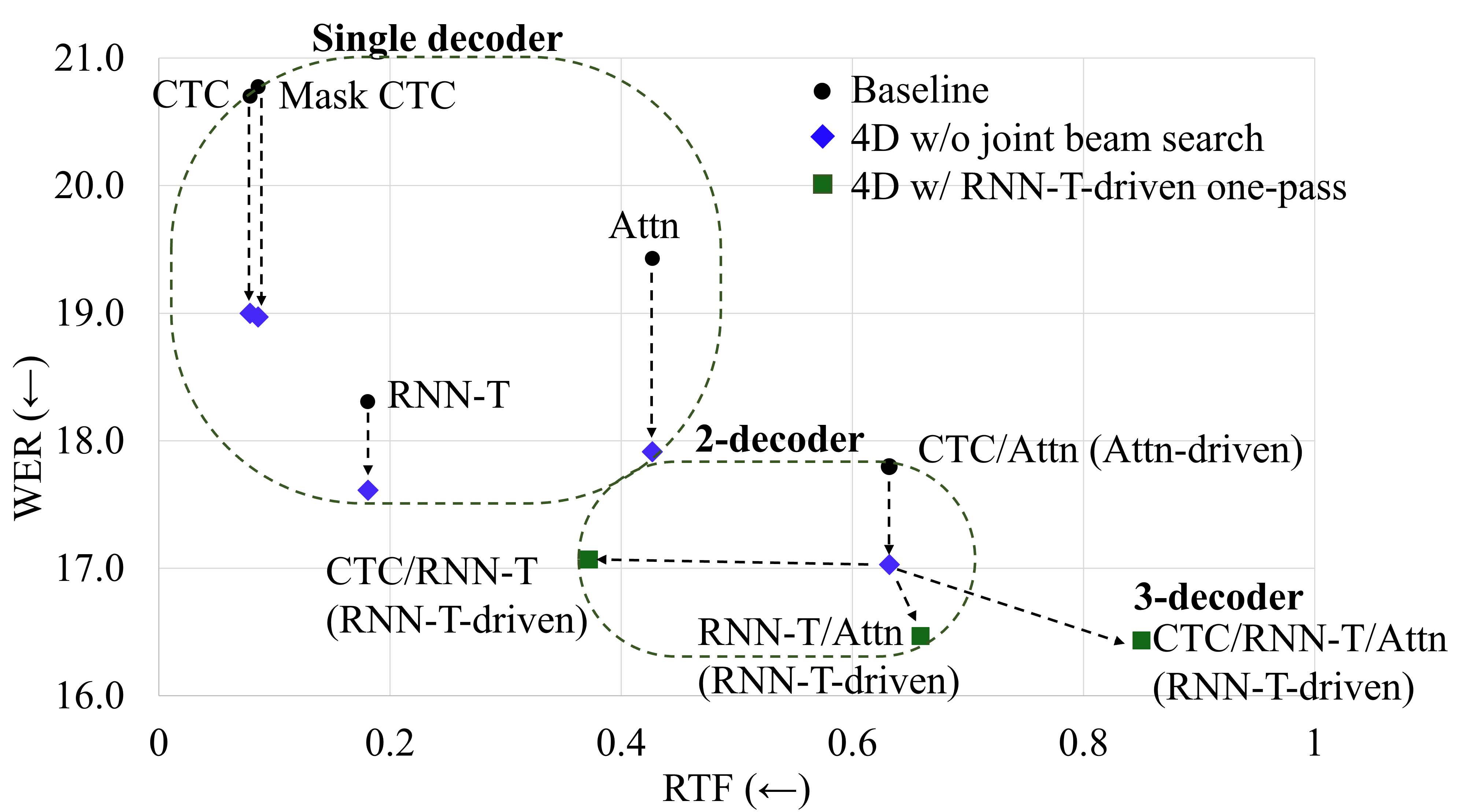} 
        \end{minipage}
        \centering
    \vspace*{-3mm}
    \caption{Relationship between RTFs and WERs. The black, blue, and green dots represent the baselines, the 4D model without beam search, and the 4D model with beam search, respectively.}
    \label{fig:rtf}
\vspace*{-2mm}
\end{figure}

Figure \ref{fig:rtf} illustrates the relationship between the real-time factor~(RTF) using a GPU (NVIDIA RTX3090) and WER on the LibriSpeech 100 hours test-other set. 
Here, the RTF is defined as follows:
\begin{equation}
    \text{RTF} = \frac{T_{\text{proc}}}{T_\text{speech}},
\label{eq:rtf}
\end{equation}
where $T_{\text{proc}}$ and $T_\text{speech}$ represent the processing time and the duration of the input speech signal, respectively. The RTF provides a measure of computational complexity in offline systems, allowing the evaluation of the processing delay relative to the input speech length.
The black dots denote the baselines, the blue dots represent 4D joint training but \textit{without} joint beam search, and the green dots denote 4D joint training \textit{with} RNN-T-driven joint beam search.
In addition to the three-decoder joint beam search of CTC/RNN-T/attention, we show the results of the two-decoder joint beam search of CTC/RNN-T and RNN-T/attention. The decoding weights ($\mu_{\text{ctc}}$, $\mu_{\text{rnnt}}$, $\mu_{\text{att}}$) of the RNN-T-driven joint beam search in Algorithm 4 are (0.3, 0.7, 0.0) and (0.0, 0.5, 0.5) for CTC/RNN-T and RNN-T/attention, respectively.

Comparing the black and blue dots, the proposed 4D model consistently reduces WER for all decoders without an increase in RTF. Although joint training requires more GPU memory and training time, as discussed in Section \ref{sec:traintime}, the proposed 4D model does not increase the decoding time as long as a single decoder is used. 

Shifting the focus to joint decoding with two decoders, compared to conventional CTC/attention joint decoding (represented by the blue dot), CTC/RNN-T achieves a smaller RTF with a comparable WER, and RNN-T/attention achieves a better WER with a similar RTF. 
In addition, the three-decoder joint beam search of CTC/RNN-T/attention shows the best WER with a 30\% larger RTF (0.85 vs. 0.65).

\begin{figure}[t!]
    \centering
        \begin{minipage}{0.49\textwidth}
            \includegraphics[width=\textwidth]{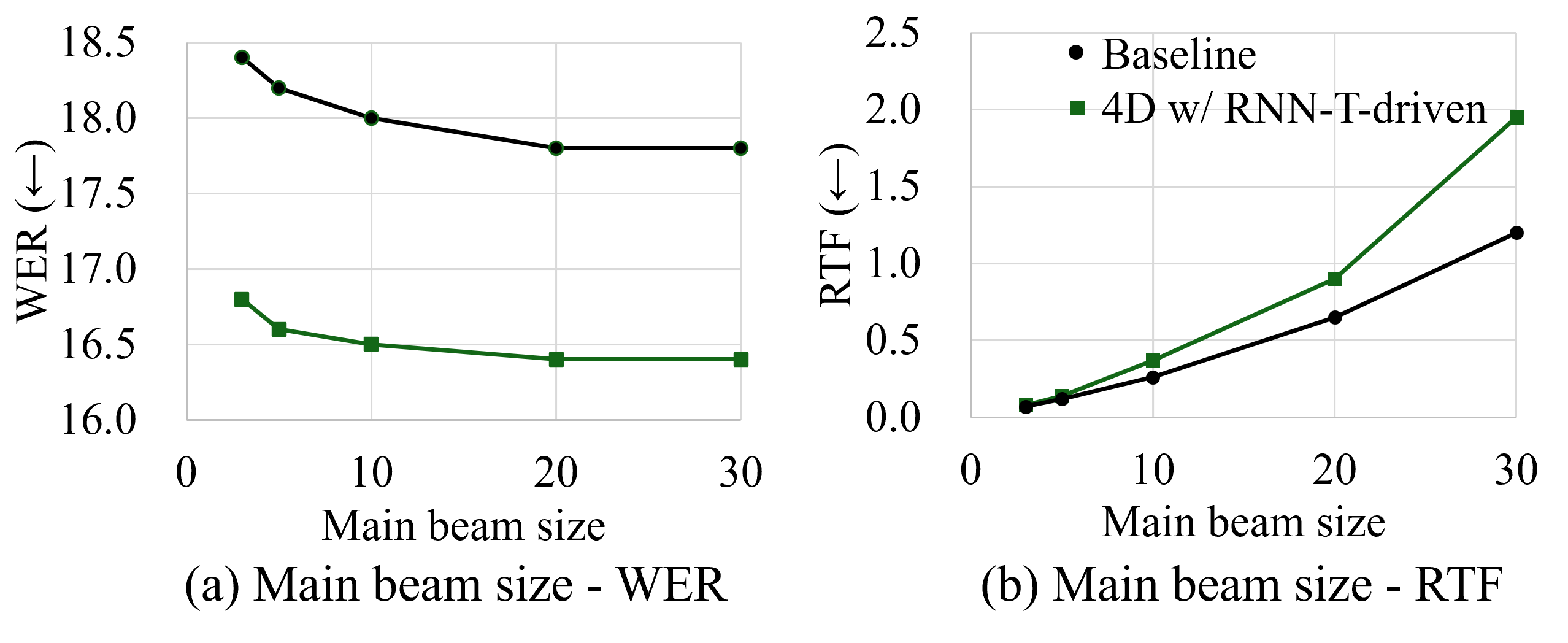} 
        \end{minipage}
        \centering
    \vspace*{-4mm}
    \caption{Relationship between main beam size and RTF/WER. The black dots represent the baselines, while the green dots indicate the 4D model with RNN-T-driven joint beam search.}
    \label{fig:beamsize_rtf}
\vspace*{-1mm}
\end{figure}

\begin{figure}[t!]
    \centering
        \begin{minipage}{0.49\textwidth}
            \includegraphics[width=\textwidth]{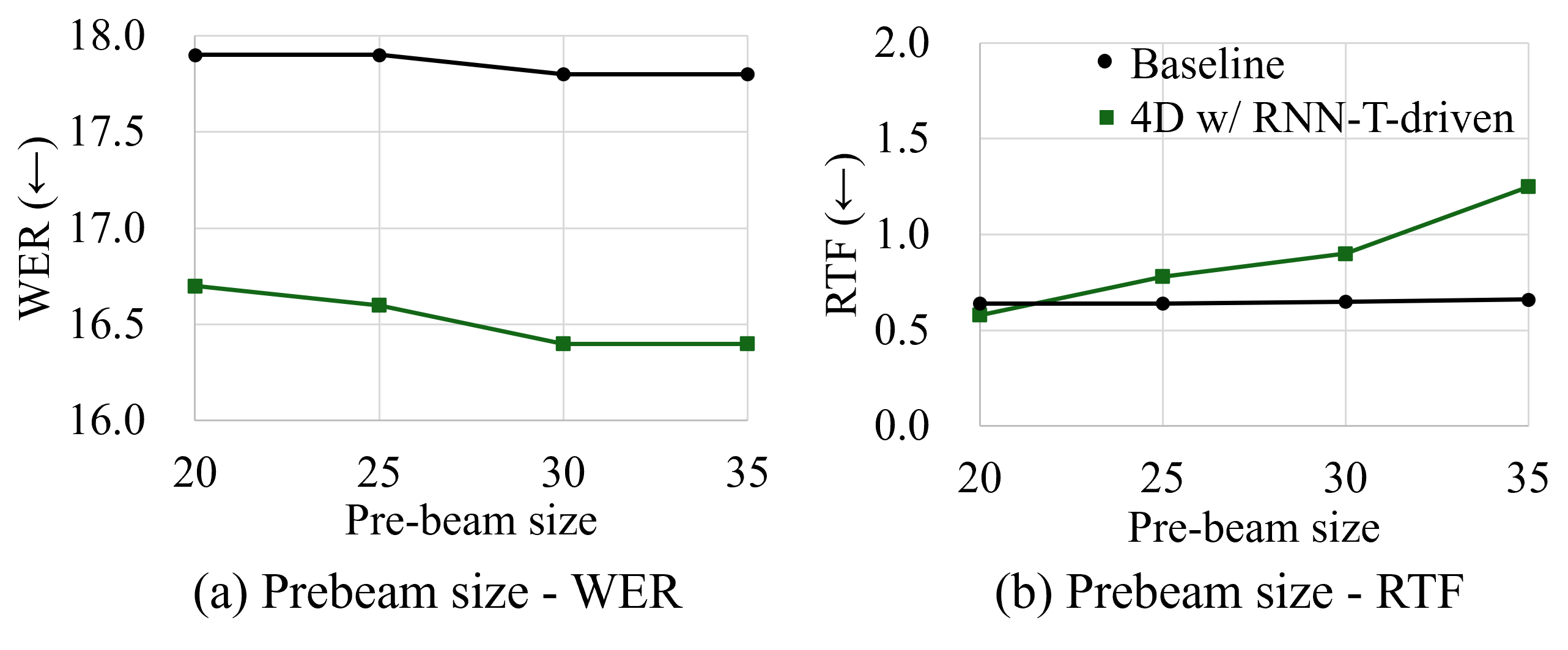} 
        \end{minipage}
        \centering
    \vspace*{-4mm}
    \caption{Relationship between prebeam size and RTF/WER when the main beam size $k_{\text{beam}}$ =  20. The black dots represent the baselines, while the green dots indicate the 4D model with RNN-T-driven joint beam search.}
    \label{fig:prebeam}
\vspace*{-1mm}
\end{figure}

Figures \ref{fig:beamsize_rtf} and \ref{fig:prebeam} demonstrate the impact of main and prebeam sizes on the performance of both the proposed method and the conventional CTC/attention joint decoding. Increasing the beam sizes improves the WER, but also increases the RTF. While the proposed method exhibits a larger increase in RTF compared to CTC/attention joint decoding, its significant WER improvements allow the RNN-T-driven joint decoding method to outperform the conventional approach, even with smaller beam sizes.

\subsubsection{Typical example of the joint decoding}
\label{sec:example}

\begin{figure}[t!]
\vspace{-0mm}
    \centering
        \begin{minipage}{0.485\textwidth}
            \includegraphics[width=\textwidth]{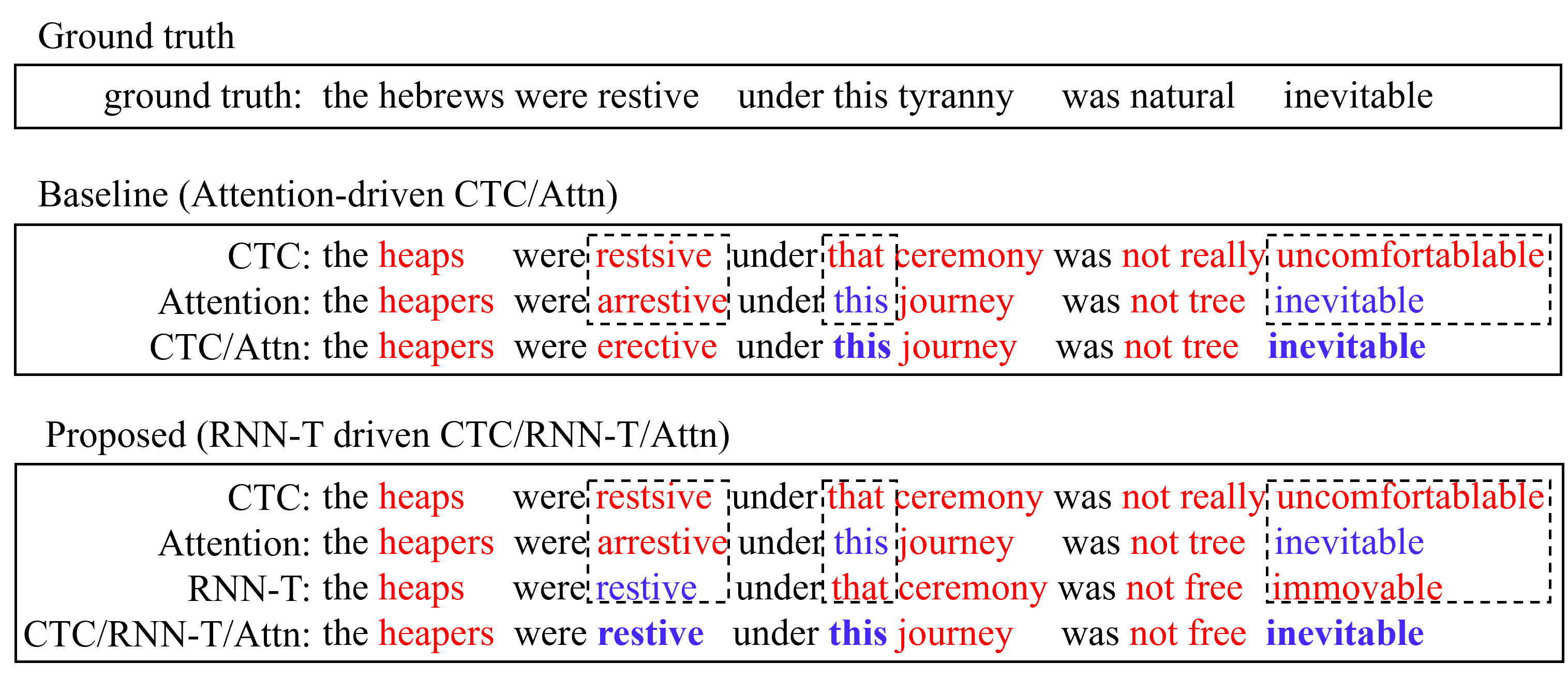} 
        \end{minipage}
        \centering
    \vspace*{-3mm}
    \caption{Typical examples of beam search results. Incorrect decoding results are highlighted in red. Conversely, correct decoding results are shown in blue, and results corrected by the joint decoding are further bolded.} 
    \label{fig:example}
\vspace*{-3.5mm}
\end{figure}

Figure \ref{fig:example} shows typical decoding results of the conventional CTC/attention and the proposed RNN-T driven joint beam search. The decoding results are also shown using each single decoder (CTC, attention, and RNN-T decoders). The incorrect decoding results are highlighted in red. 
Conversely, the correct decoding results are shown in blue, and results corrected by the joint decoding are further bolded.
In both CTC/attention and the proposed joint beam search, joint decoding can correct the incorrectly decoded result if any decoder outputs a correct token.
The proposed method integrates the three decoders, thereby enhancing the complementarity of each decoder resulting in a performance improvement in terms of WER.

\subsubsection{Effect of the decoding weights}
\label{sec:decodeweight}

\begin{figure}[t!]
    \centering
        \begin{minipage}{0.45\textwidth}
            \includegraphics[width=\textwidth]{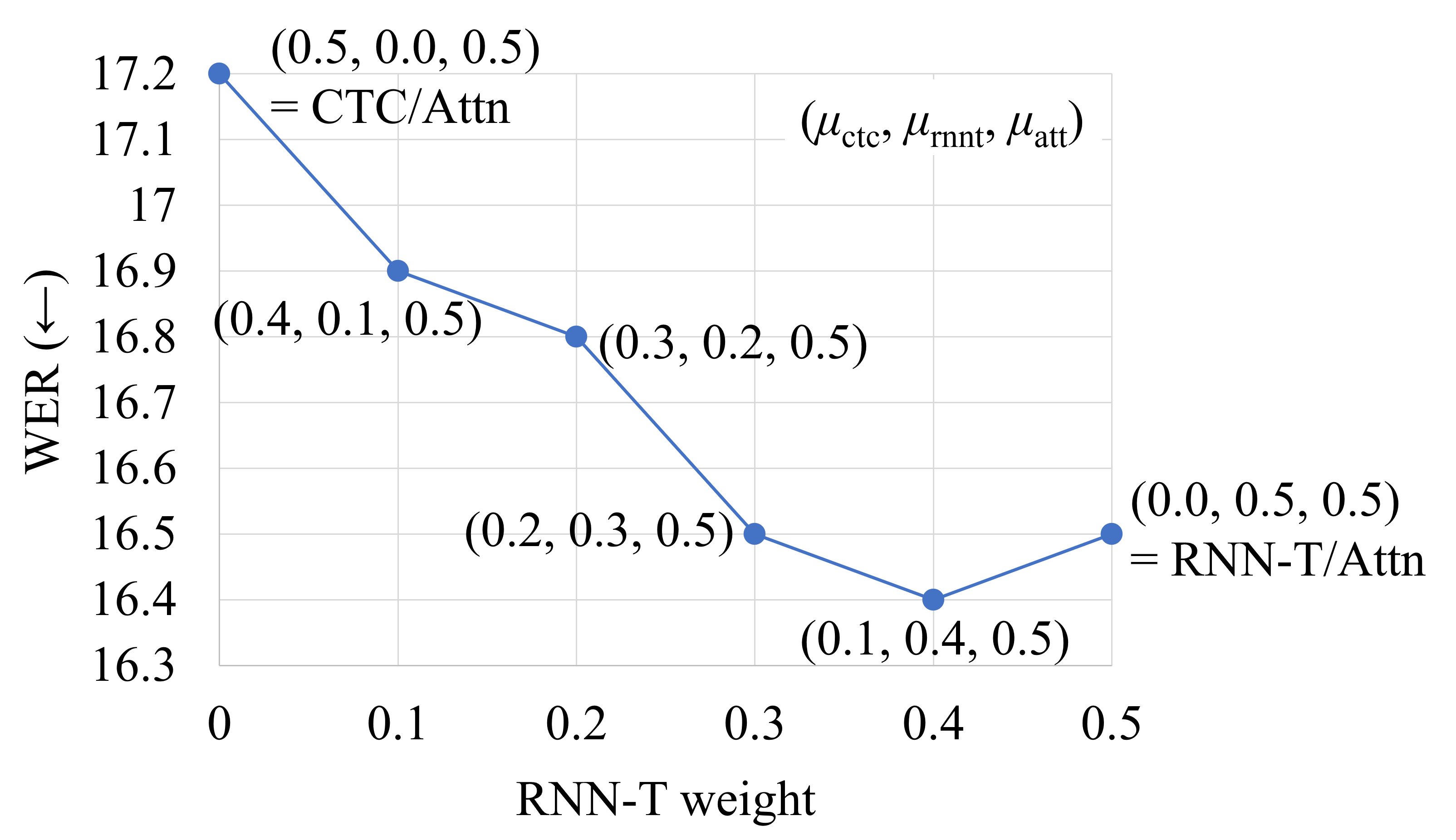} 
        \end{minipage}
    \vspace*{-0mm}
    \caption{Effect of the decoding weights. We tested the impact of the RNN-T weights by fixing the attention weight to 0.5 for CTC/attention. The proposed method consistently outperformed the baseline by increasing the RNN-T weight.} 
    \label{fig:decodeweight}
\vspace*{-1mm}
\end{figure}

As depicted in Figure~\ref{fig:decodeweight}, we examine the effect of RNN-T weight $\mu_{\text{rnnt}}$ by maintaining the attention weight $\mu_{\text{att}}$ at 0.5 for CTC/attention. The proposed method consistently outperforms the baseline by increasing the RNN-T weight, with the best performance observed when utilizing the decoding weights ($\mu_{\text{ctc}}$, $\mu_{\text{rnnt}}$, $\mu_{\text{att}}$) =  (0.1, 0.4, 0.5). 

\subsection{Comprehensive analysis of the joint beam search}
\label{sec:decodeanalisys}

This section provides a comprehensive analysis of the three proposed joint beam search algorithms in terms of WER-RTF tradeoff, decoder combination, and comparison with another joint decoding method.

\subsubsection{Comparison of the three joint beam search}
\label{sec:comparison}

\begin{figure*}[t!]
    \centering
        \begin{minipage}{1.0\textwidth}
            \includegraphics[width=\textwidth]{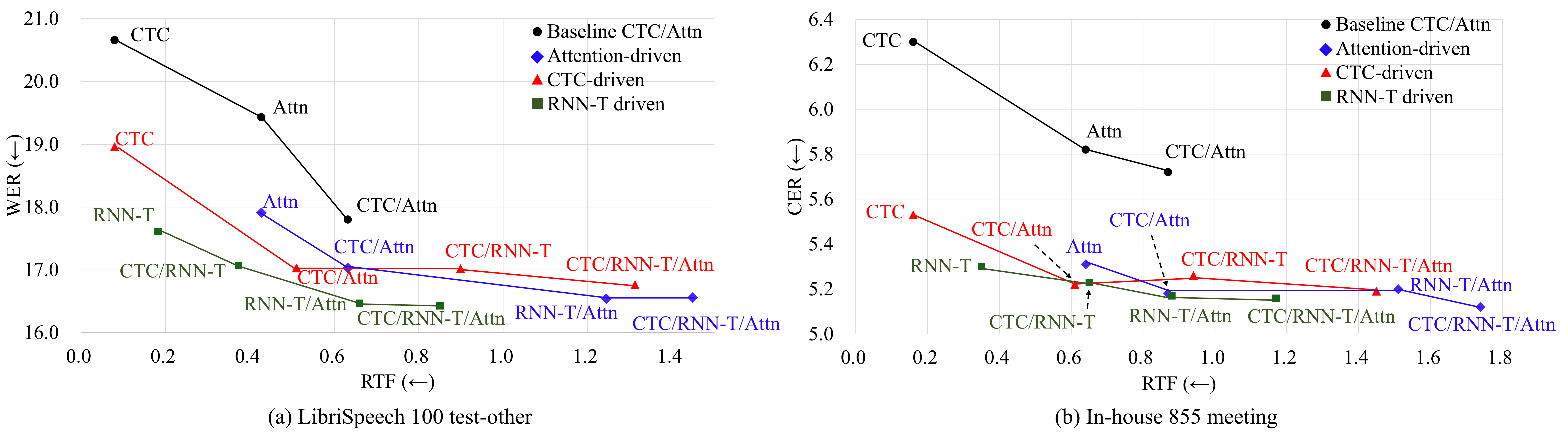} 
        \end{minipage}
        \centering
    \vspace*{-4mm}
    \caption{Comparison of the three joint beam search methods. The black dots represent the conventional CTC/attention, while the red, blue, and green dots depict the results of the attention-driven, CTC-driven, and RNN-T-driven joint beam search results, respectively.} 
    \label{fig:comparison}
\vspace*{-4mm}
\end{figure*}

Figure \ref{fig:comparison} illustrates the relationship between the RTF using a GPU (NVIDIA RTX3090) and WER on the LibriSpeech 100 hours test-other set and our in-house dataset. The black dots represent the baseline CTC/attention results, while the blue, red, and green dots represent the attention-driven, CTC-driven, and RNN-T-driven joint beam search, respectively. We also show the results for single-decoder and two-decoder joint beam search.
The decoding weights ($\mu_{\text{ctc}}$, $\mu_{\text{rnnt}}$, $\mu_{\text{att}}$) for the two-decoder joint beam search in Algorithms 1, 3, and 4 are (0.3, 0.7, 0.0), (0.3, 0.0, 0.7), and (0.0, 0.5, 0.5) for CTC/RNN-T, CTC/attention, and RNN-T/attention, respectively.

Upon comparison of the baseline with the three proposed joint beam search algorithms, all three methods exhibit superior WER-RTF tradeoff curves, reflecting the improved baseline performance achieved through joint training.
Notably, even with the same decoder combinations, differences in the WER-RTF tradeoff can be observed. For example, the CTC-driven CTC/attention joint decoding method achieves a lower RTF than the attention-driven method (red and blue dots in Figure \ref{fig:comparison}), while maintaining a comparable WER. This finding is consistent with the results reported in \cite{yan2022ctc} for speech translation tasks.
Among the three proposed joint beam search algorithms, the RNN-T-driven method achieves the best WER-RTF tradeoff.
This advantage stems from the fact that the CTC-driven and attention-driven methods compute all RNN-T paths through RNN-T prefix scoring, whereas the RNN-T-driven joint beam search avoids this computation.

\begin{figure}[t!]
    \centering
        \begin{minipage}{0.49\textwidth}
            \includegraphics[width=\textwidth]{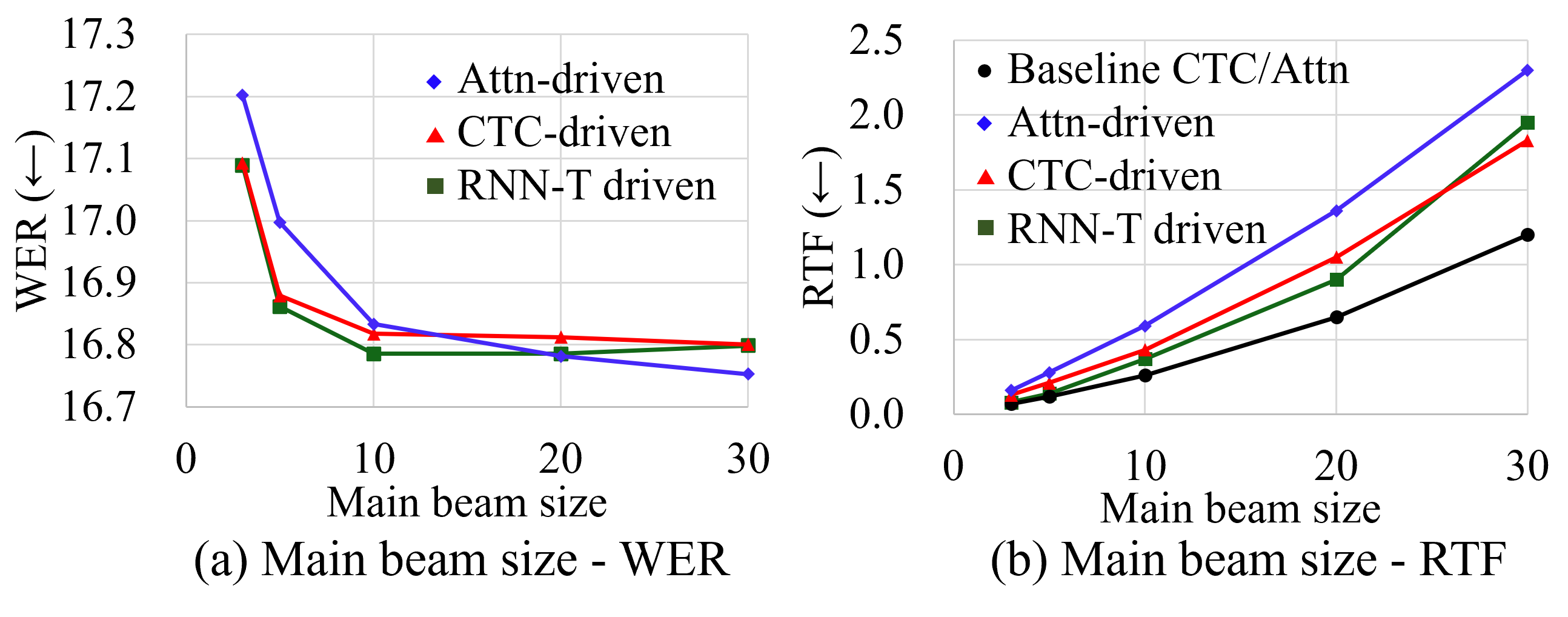} 
        \end{minipage}
        \centering
    \vspace*{-5mm}
    \caption{Relationship between \textcolor{black}{main beam} size and RTF, WER. The black, red, blue, and green dots represent the baselines and the 4D model with the joint beam search, respectively.}
    \label{fig:beamsize_cpmarison}
\vspace*{-0mm}
\end{figure}

When the three decoders are integrated, as shown in Figure \ref{fig:comparison} and Table \ref{maintable} (B7-9), different WERs are observed depending on the primary decoder, even though the same decoders are integrated. Figure \ref{fig:beamsize_cpmarison} (a) shows the differences in WER reduction trends depending on the primary decoder, with the decoder weights set to ($\mu_{\text{ctc}}$, $\mu_{\text{rnnt}}$, $\mu_{\text{att}}$) = (0.33, 0.33, 0.34).
Figure \ref{fig:beamsize_cpmarison} (a) shows that, when the same decoders are integrated with the same decoder weights, the WER results still differ. This variation is due to the accuracy of the hypotheses generated by each decoder. 
Specifically, as discussed in Section \ref{sec:Preliminary}, the attention decoder tends to generate more misaligned hypotheses, whereas the CTC and RNN-T decoders, which assume monotonic alignment, generate more accurate hypotheses with smaller beam sizes. Among these, the RNN-T decoder generates the most accurate hypotheses because it relaxes the conditional independence assumption of the CTC decoder. 
Although increasing the beam size will eventually result in all three methods generating the same hypotheses, this approach is impractical due to the significant increase in RTF (Figure \ref{fig:beamsize_cpmarison} (b)).

\begin{figure}[t!]
    \centering
        \begin{minipage}{0.49\textwidth}
            \includegraphics[width=\textwidth]{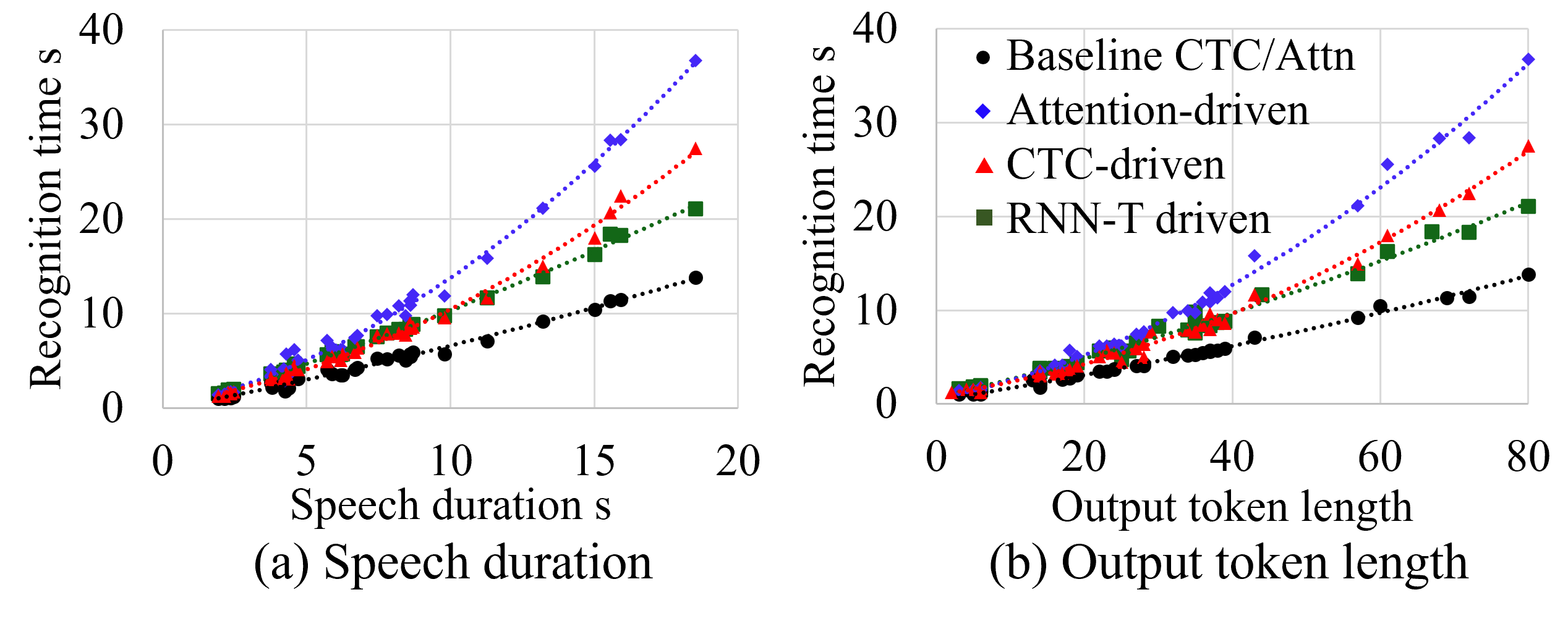} 
        \end{minipage}
        \centering
    \vspace*{-4mm}
    \caption{Relationship between input/output length and recognition time. The black, red, blue, and green dots represent the baselines and the 4D model with the joint beam search, respectively.}
    \label{fig:duration}
\vspace*{-1mm}
\end{figure}

Figure \ref{fig:duration} shows the relationship between recognition time and input/output length. Consistent with the results in Figure \ref{fig:comparison}, the recognition time for the proposed joint decoding methods is longer than for conventional CTC/attention due to the integration of three decoders. However, among the proposed methods, RNN-T-driven joint decoding has the shortest recognition time.

\subsubsection{Analysis of different decoder combinations}
\label{sec:combination}

\begin{figure}[t!]
\vspace{-0mm}
    \centering
        \begin{minipage}{0.49\textwidth}
            \includegraphics[width=\textwidth]{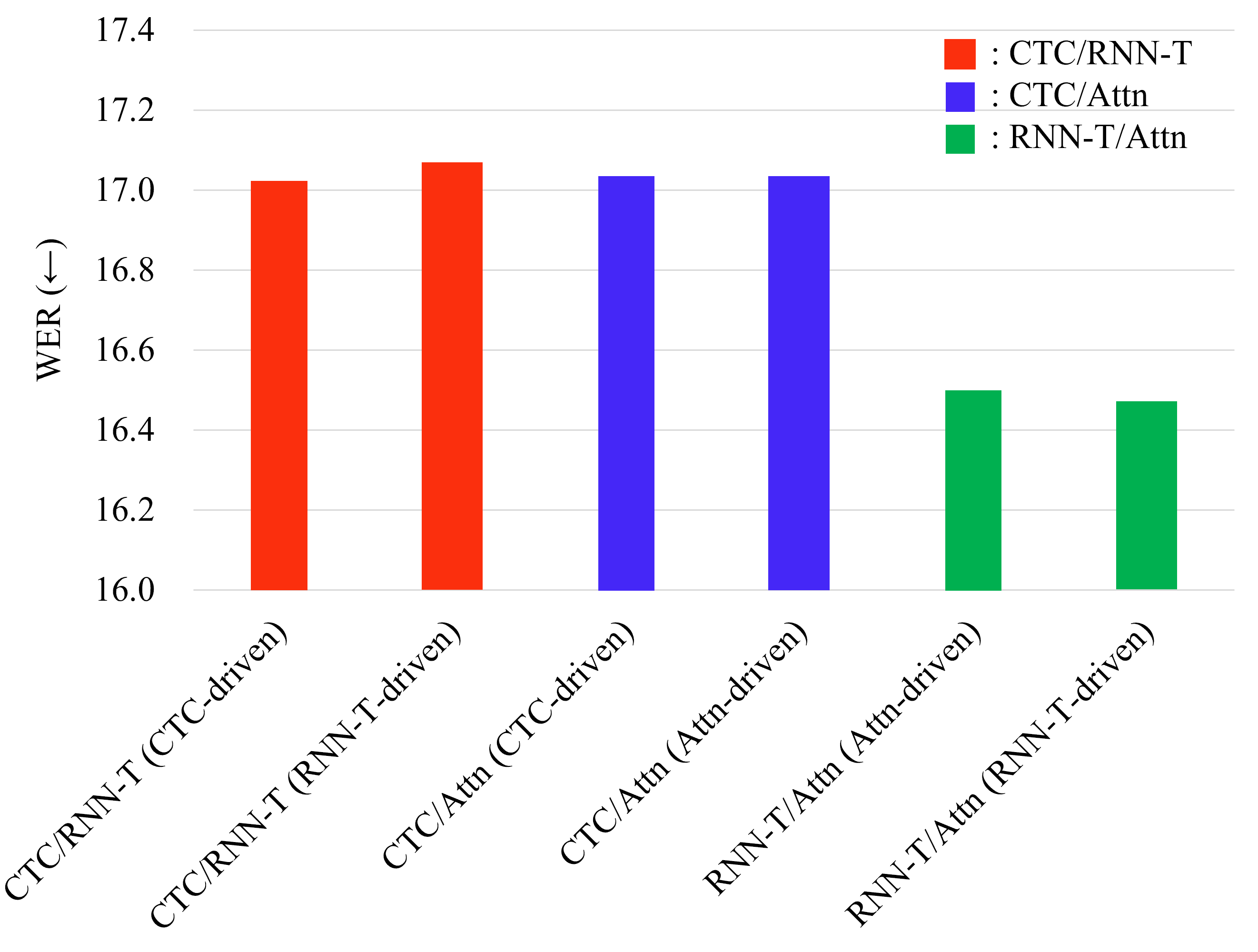} 
        \end{minipage}
        \centering
    \vspace*{-4.5mm}
    \caption{Comparison of performance between different combinations of beam search strategies.}
    \label{fig:combination}
\vspace*{-3mm}
\end{figure}

Figure \ref{fig:combination} shows the performance when using two of the CTC, RNN-T, and attention decoders (CTC/RNN-T, CTC/attention, RNN-T/attention). A total of six different combinations are compared, depending on which decoder is designated as the primary decoder.
Interestingly, regardless of the primary decoder, when the decoder used for joint scoring is the same, the performance is nearly comparable.
Within this experimental setting, the RNN-T/attention combination exhibits the lowest WER. This result is primarily attributed to the inherent performance capabilities of each decoder. 
Considering the WER-RTF tradeoff shown in Figure \ref{fig:comparison}, the proposed method can switch between decoders to show reasonable performance with smaller RTFs for different application scenarios.

\subsubsection{Comparison of performance with ROVER}
\label{sec:rover}

In addition, we compare the proposed joint beam search method with the recognizer output voting error reduction method (ROVER) \cite{fiscus1997post} in Table \ref{rover}.
Overall, ROVER and the three proposed joint beam search 
 algorithms show almost equal WER, but the RNN-T-driven joint beam search shows the best WER.
Although ROVER is a powerful method, it increases system complexity and computation time because it requires complete token sequences from all decoders for integration.
In contrast, the proposed method is more efficient because the secondary decoders do not need to generate complete token sequences.

\begin{table}[t]
\caption{WER ($\downarrow$) comparison: joint beam search vs. ROVER on LibriSpeech test-(clean/other) set.}
\vspace*{-8mm}
\label{rover}
\begin{center}
\resizebox {\linewidth} {!} {
\begin{tabular}{@{}l|cc|cc}
\hline
& \multicolumn{2}{c|}{LibriSpeech 960} & \multicolumn{2}{c}{LibriSpeech 100} \\
 Decoding method & clean & other & clean & other\\
\hline
Attn-driven CTC/RNN-T/Attn & \textbf{2.4} & 5.3 & 6.5 & 16.6 \\
CTC-driven CTC/RNN-T/Attn  & \textbf{2.4} & 5.3 & \textbf{6.3} & 16.8\\ 
RNN-T-driven CTC/RNN-T/Attn & \textbf{2.4} & \textbf{5.2} & \textbf{6.3} & \textbf{16.4} \\
\hline
ROVER (CTC/RNN-T/Attn) & \textbf{2.4} & 5.3 & \textbf{6.3} & 16.6\\
ROVER (CTC/RNN-T/Attn/Mask-CTC) & \textbf{2.4} & 5.3 & 6.4 & 16.8 \\
\hline
\end{tabular}
}
\end{center}
\vspace*{-4mm}
\end{table}

\section{Discussion}
\label{sec:discussion}

Summarizing the findings from Section \ref{sec:experiments}, several key observations emerge: 1) performance improves with an increasing number of decoders used for joint training (Section \ref{sec:3d4d}). 2) A larger number of decoders employed for joint decoding enhances performance, as reflected in the WER-RTF tradeoff (Section \ref{sec:rtf-wer}). 3) Notably, the RNN-T-driven joint beam search demonstrates a superior WER-RTF tradeoff compared to other joint decoding methods, including CTC-driven and attention-driven as well as the baseline CTC/attention curves (Section \ref{sec:decodeanalisys}).

An inherent advantage of the 4D model lies in its ability to switch between decoders depending on the application scenario. Despite the increased training and decoding time associated with joint training and decoding, this model eliminates the need for training a different model for each application. 
Once the joint training is complete, the 4D model provides a unified framework that can be easily adapted to different application scenarios by simply switching between decoders. 

In addition, depending on the application scenario, the 4D model allows for the selection of an appropriate decoding algorithm, such as a two-decoder or three-decoder joint beam search. 
For example, the CTC decoder is appropriate for real-time processing with limited computational resources. In contrast, the joint beam search with three decoders becomes viable when sufficient computational resources are available.
By offering this flexibility, the 4D model not only improves performance but also optimizes resource usage during deployment across different environments.

Furthermore, the proposed method can be extended to other sequence-to-sequence tasks, such as speech and machine translation. Although the monotonic alignment constraints of CTC and RNN-T decoders pose challenges for non-monotonic alignment tasks, the hierarchical encoder structure proposed in \cite{yan2022ctc} enables these decoders to address such tasks. This highlights the versatility of the 4D model across various sequence-to-sequence scenarios, enhancing its practical applicability.

Considering these attributes, the proposed 4D model is a practical system capable of achieving a superior WER-RTF tradeoff curve compared to the baseline CTC/attention.

\section{Conclusion}
\label{sec:conclusion}

This paper introduces a 4D joint model incorporating CTC, attention, RNN-T, and Mask-CTC, with a shared encoder trained through joint training. 
The study demonstrates the efficacy of jointly trained 4D models, employing a two-stage training strategy that enhances the performance of individual decoders. 
Notably, the proposed joint CTC/RNN-T/attention decoding exhibits improved performance, surpassing the previously proposed CTC/attention decoding. 
The RNN-T-driven joint beam search, in particular, showcases a superior WER-RTF tradeoff compared to other joint decoding methods, including CTC-driven, attention-driven joint beam search algorithms, and the baseline CTC/attention curves.


\bibliographystyle{IEEEtran}
\bibliography{mybib}


 
\vspace{11pt}

\begin{IEEEbiography}[{\includegraphics[width=1in,height=1.25in,clip,keepaspectratio]{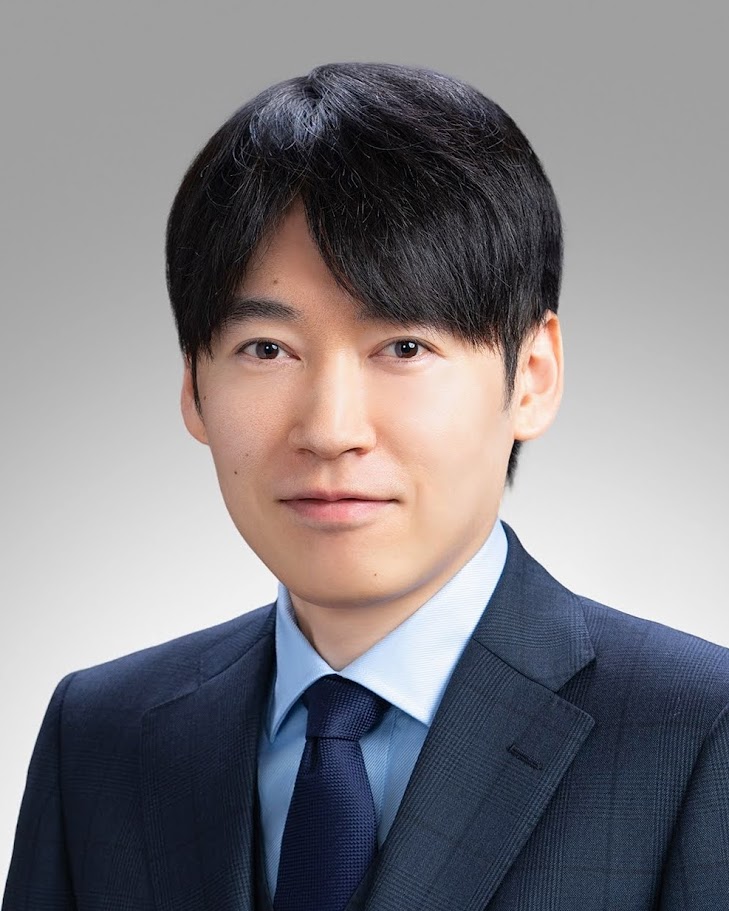}}]{Yui Sudo}
received the B.S. and M.S. degrees from Keio University, Japan, in 2009 and 2011, respectively, and the Ph.D. degree from Tokyo Institute of Technology, Japan, in 2021. He worked with Honda Motor Co., Ltd., Japan, in 2011, Honda Engineering Co., Ltd., Japan, from 2012 to 2018, and Honda R\&D Co., Ltd., Japan, from 2019 to 2020. Currently, he is a Senior Engineer at Honda Research Institute Japan Co., Ltd., Japan. His research interests include automatic speech recognition, computational auditory scene analysis, and robot audition. 
\end{IEEEbiography}

\begin{IEEEbiography}[{\includegraphics[width=1in,height=1.25in,clip,keepaspectratio]{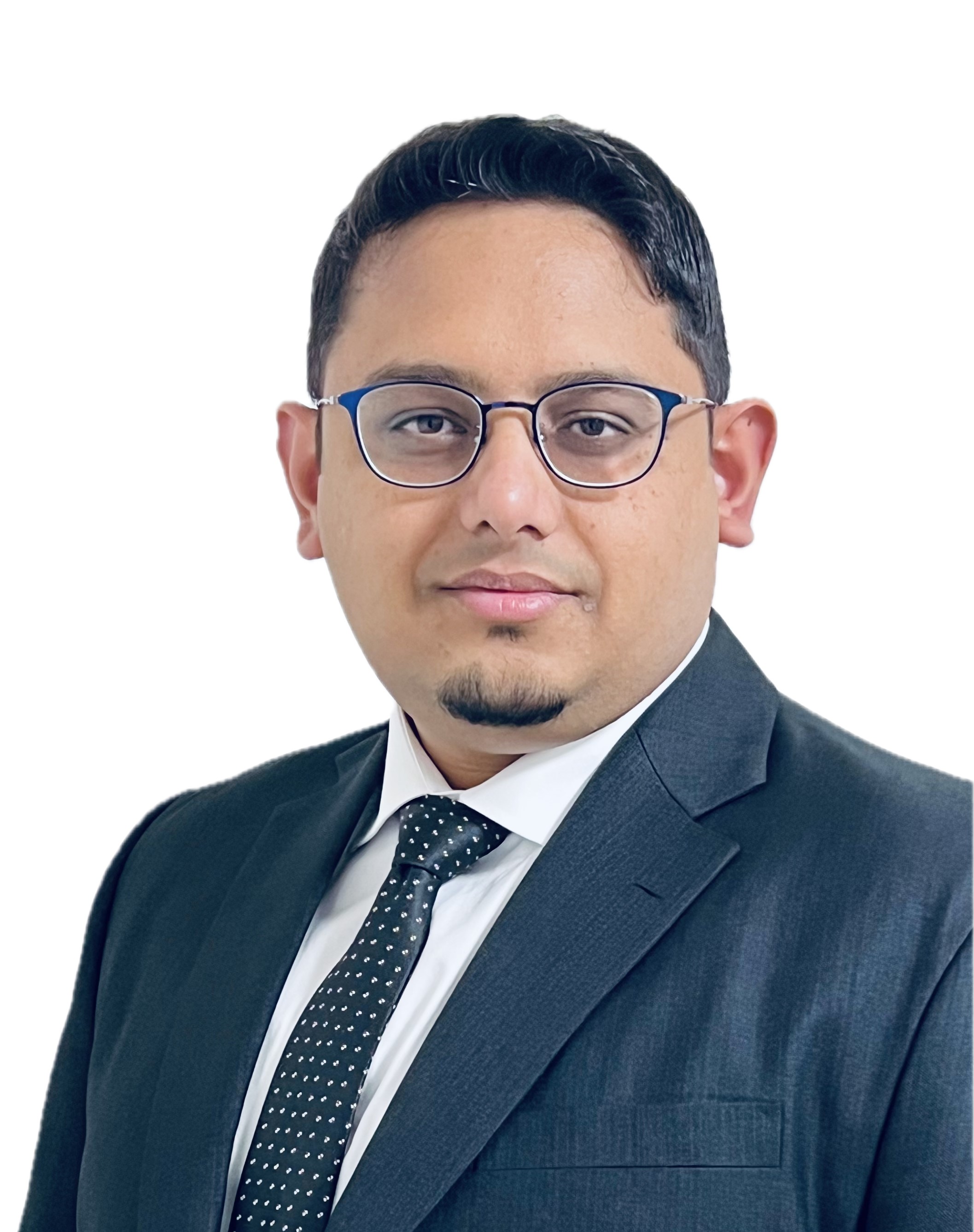}}]{Muhammad Shakeel}
received a B.S. in Electronics from COMSATS University Islamabad in 2010, an M.S. in Artificial Intelligence and Robotics as an Erasmus Mundus scholar from the Sapienza University of Rome in 2015, and a Ph.D. in Systems and Control Engineering as a MEXT scholar from the Institute of Science Tokyo (former Tokyo Institute of Technology) in 2022. He worked at COMSATS University Islamabad as a Lecturer for three years, from 2016 to 2019. Currently, he is a Scientist at Honda Research Institute Japan Co., Ltd. His research interests include artificial intelligence, automatic speech recognition, robot audition, and multimodal integration.
\end{IEEEbiography}

\begin{IEEEbiography}[{\includegraphics[width=1in,height=1.25in,clip,keepaspectratio]{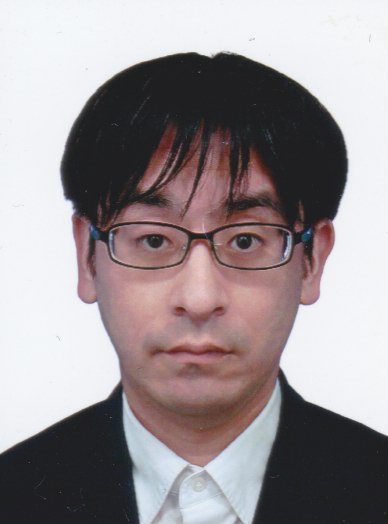}}]{Yosuke Fukumoto}
received the B.S. and M.S. degrees from Osaka University, Japan. Currently, he is a Guest Engineer at Honda Research Institute Japan Co., Ltd., Saitama, Japan. His research interests include automatic speech recognition, reinforcement learning, and artificial intelligence.
\end{IEEEbiography}

\begin{IEEEbiography}[{\includegraphics[width=1in,height=1.25in,clip,keepaspectratio]{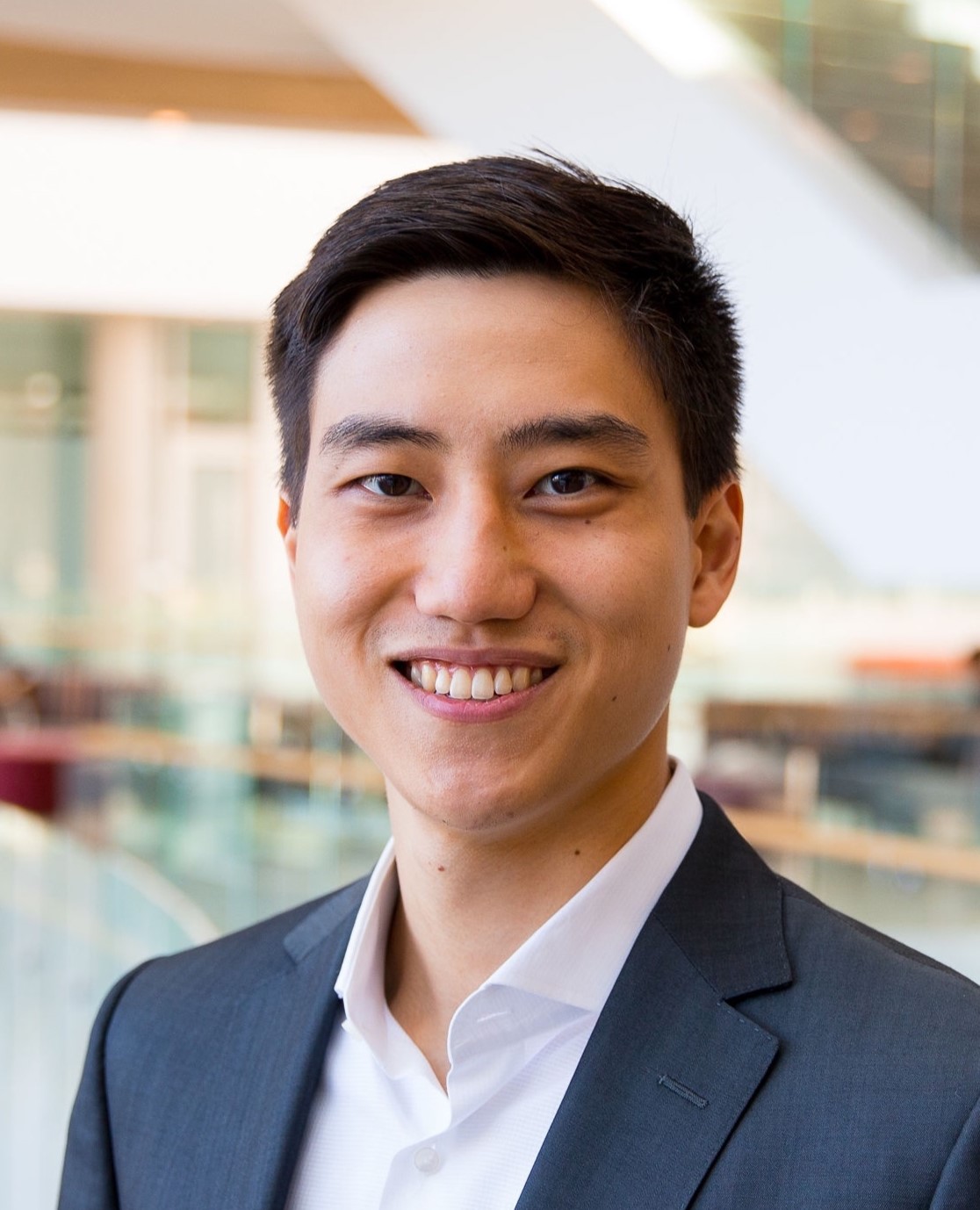}}]{Brian Yan} received a M.S. in Computer Science from Carnegie Mellon University in 2021 and a B.A. in Economics from University of Chicago in 2016. He is currently a Ph.D. student in the Language Technologies Institute at Carnegie Mellon University. His current research focuses on multi-sequence transduction for speech applications such as speech-to-text translation and multilingual speech recognition.
\end{IEEEbiography}

\begin{IEEEbiography}[{\includegraphics[width=1in,height=1.25in,clip,keepaspectratio]{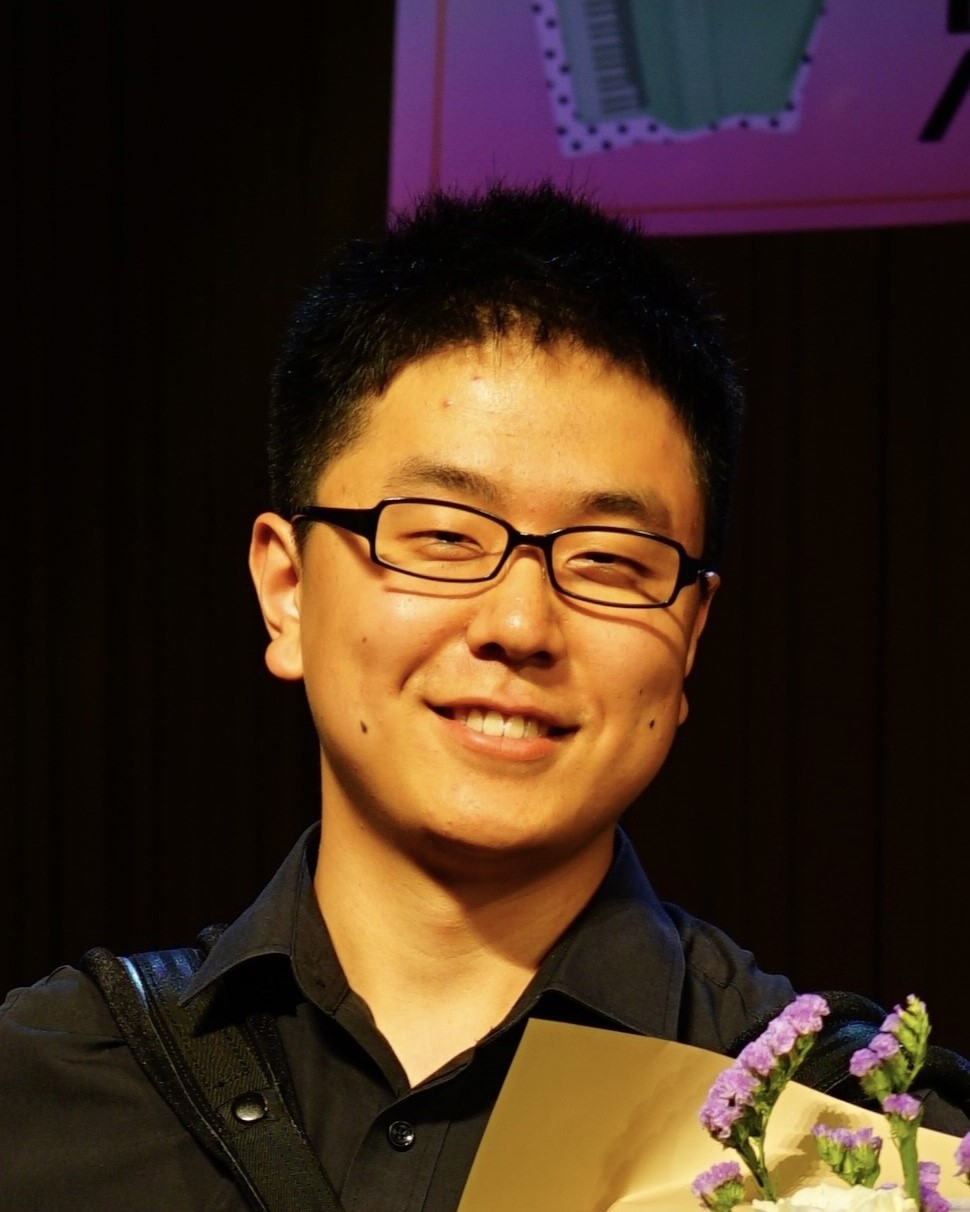}}]{Jiatong Shi}
is a Ph.D. student, advised by Dr. Shinji Watanabe, at Watanabe’s Audio and Voice (WAV) Lab, Language Technologies Institute, Carnegie Mellon University (CMU). He received his B.S. in computer science from Renmin University of China (RUC), advised by Dr. Qin Jin, and M.S. in computer science from Johns Hopkins University (JHU) advised by Dr. Shinji Watanabe. His major research focus is speech representation learning and its applications, including speech recognition, speech translation, speech, and singing voice synthesis.
\end{IEEEbiography}

\begin{IEEEbiography}[{\includegraphics[width=1in,height=1.25in,clip,keepaspectratio]{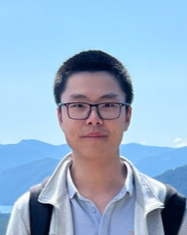}}]{Yifan Peng}
is a Ph.D. candidate in Electrical and Computer Engineering at Carnegie Mellon University, PA, USA, advised by Prof. Shinji Watanabe. He received a B.E. from Tsinghua University, Beijing, China, in 2020. His research focuses on developing effective and efficient speech foundation models for various speech-processing tasks, such as speech recognition and understanding.
\end{IEEEbiography}

\begin{IEEEbiography}[{\includegraphics[width=1in,height=1.25in,clip,keepaspectratio]{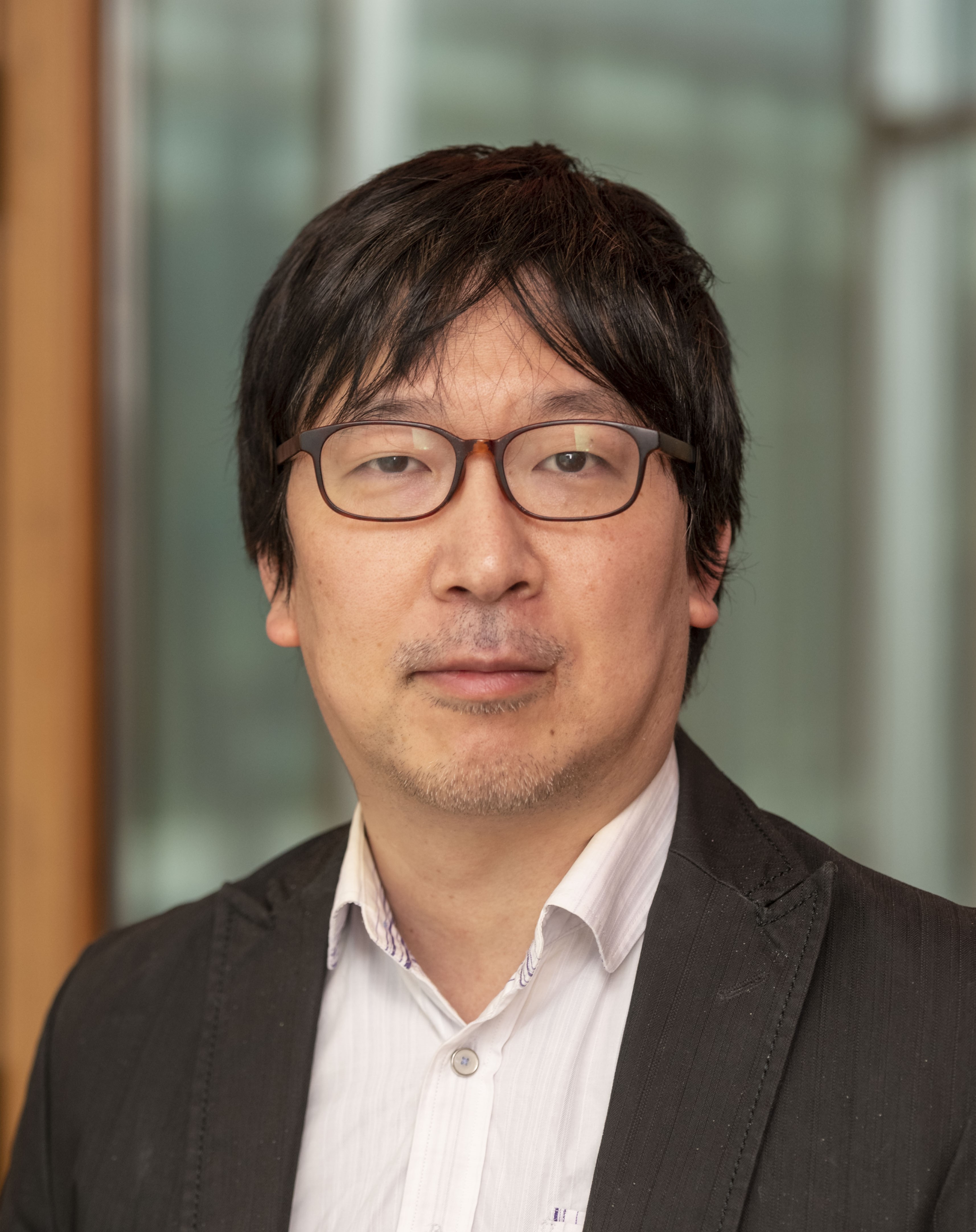}}]{Shinji Watanabe}
received the B.S., M.S., and Ph.D. (Dr. Eng.) degrees from Waseda University, Tokyo, Japan. He was a Research Scientist with NTT Communication Science Laboratories, Kyoto, Japan, from 2001 to 2011, a visiting scholar with Georgia Institute of Technology, Atlanta, GA, USA, in 2009, and a Senior Principal Research Scientist with Mitsubishi Electric Research Laboratories (MERL), Cambridge, MA, USA, from 2012 to 2017. Before Carnegie Mellon University, he was an Associate Research Professor with Johns Hopkins University, Baltimore, MD, USA, from 2017 to 2020. He is currently an Associate Professor with Carnegie Mellon University, Pittsburgh, PA, USA. He has authored or co-authored more than 500 papers in peer-reviewed journals and conferences and received several awards, including the best paper award from the IEEE ASRU in 2019. His research interests include automatic speech recognition, speech enhancement, spoken language understanding, and machine learning for speech and language processing. He is a Senior Area Editor of IEEE Transactions on Audio, Speech, and Language Processing. He was/has been a member of several technical committees, including the APSIPA Speech, Language, and Audio Technical Committee (SLA), IEEE Signal Processing Society Speech and Language Technical Committee (SLTC chair in 2025--2026), and Machine Learning for Signal Processing Technical Committee (MLSP). He is an ISCA Fellow.
\end{IEEEbiography}



\vfill

\end{document}

%% file: main_table.tex
\resizebox {1.00\linewidth} {!} {

\begin{tabular}{llccc|cc|cc|cc|c}
\hline

 & & 4D joint & Primary & Params M & \multicolumn{2}{c|}{LibriSpeech 960 (WER)} & \multicolumn{2}{c|}{LibriSpeech 100 (WER)} & \multicolumn{2}{c|}{In-house 855 (CER)} & Avg \\
ID & Model Name & training & decoder & (decode/train) & test-clean & test-other & test-clean & test-other & assembly & meeting & $\Delta$ \\

\hline

A1 & Attention & - & - & 113.6/116.2 & 2.9 & 5.6 & 8.6 & 19.4 & 3.8 & 5.8 & - \\
B1 & 4D (Attention) & yes & - & 113.6/155.0 & \textbf{2.7} & 5.6 & \textbf{7.8} & \textbf{17.9} & \underline{\textbf{3.6}} & \textbf{5.3} & -0.5\\

\hline
A2 & CTC & - & - & 85.9/116.2 & 3.1 & 6.9 & 8.2 & 20.7 & 3.9 & 6.3 & -\\
B2 & 4D (CTC) & yes & - & 85.9/155.0 & \textbf{2.8} & \textbf{6.4} & \textbf{7.3} & \textbf{19.0} & \textbf{3.7} & \textbf{5.5} & -0.7 \\

\hline

A3 & Mask-CTC & - & - & 116.2/116.2 & \textbf{3.0} & 6.9 & 8.7 & 20.8 & 4.4 & 6.6 & -\\
B3 & 4D (Mask-CTC) & yes & - & 116.2/155.0 & 3.1 & \textbf{6.8} & \textbf{7.5} & \textbf{19.0} & \textbf{3.9} & \textbf{6.2} & -0.6\\

\hline

A4 & RNN-T & - & - & 91.8/91.8 & 2.7 & 5.8 & 7.3 & 18.3 & 4.0 & 5.9 & - \\
B4 & 4D (RNN-T) & yes & - &91.8/155.0 & \textbf{2.6} & \textbf{5.7} & \textbf{7.1} & \textbf{17.6} & \textbf{3.9} & \textbf{5.3} & -0.3 \\

\hline

A5 & CTC/Attn (2-pass) \cite{yao2021wenet} & - & CTC & 116.2/116.2 & \underline{\textbf{2.4}} & \underline{\textbf{5.2}} & 6.9 & 18.2 & 3.8 & 5.8 & - \\
B5 & 4D (CTC/Attn) & yes & CTC & 116.2/155.0 & \underline{\textbf{2.4}} & 5.3 & \textbf{6.5} & \textbf{17.0} & \textbf{3.7} & \textbf{5.2} & -0.4 \\

\hline

A6 & CTC/Attn \cite{watanabe2017hybrid} & - & Attn & 116.2/116.2 & 2.5 & \underline{\textbf{5.2}} & 7.1 & 17.8 & 3.8 & 5.7 & - \\
B6 & 4D (CTC/Attn) & yes & Attn & 116.2/155.0 & \underline{\textbf{2.4}} & 5.3 & \textbf{6.5} & \textbf{17.0} & \textbf{3.7} & \textbf{5.2} & -0.3 \\

\hline

A6 & CTC/Attn \cite{watanabe2017hybrid} & - & Attn & 116.2/116.2 & 2.5 & \underline{\textbf{5.2}} & 7.1 & 17.8 & 3.8 & 5.7 & - \\

B7 & 4D (CTC/RNN-T/Attn) & yes & Attn & \textcolor{black}{124.7}/155.0 & \underline{\textbf{2.4}} & 5.3 & \textbf{6.5} & \textbf{16.6} & \textbf{3.7} & \underline{\textbf{5.1}} & -0.4 \\
B8 & 4D (CTC/RNN-T/Attn) & yes & CTC & \textcolor{black}{124.7}/155.0 & \underline{\textbf{2.4}} & 5.3 & \underline{\textbf{6.3}} & \textbf{16.8} & \textbf{3.7} & \underline{\textbf{5.1}} & -0.4\\
B9 & 4D (CTC/RNN-T/Attn) & yes & RNN-T & \textcolor{black}{124.7}/155.0 & \underline{\textbf{2.4}} & \underline{\textbf{5.2}} & \underline{\textbf{6.3}} & \underline{\textbf{16.4}} & \textbf{3.7} & \textbf{5.2} & -0.5 \\
\hline
\end{tabular}
}

%% file: main.bbl
\begin{thebibliography}{10}
\providecommand{\url}[1]{#1}
\csname url@samestyle\endcsname
\providecommand{\newblock}{\relax}
\providecommand{\bibinfo}[2]{#2}
\providecommand{\BIBentrySTDinterwordspacing}{\spaceskip=0pt\relax}
\providecommand{\BIBentryALTinterwordstretchfactor}{4}
\providecommand{\BIBentryALTinterwordspacing}{\spaceskip=\fontdimen2\font plus
\BIBentryALTinterwordstretchfactor\fontdimen3\font minus
  \fontdimen4\font\relax}
\providecommand{\BIBforeignlanguage}[2]{{%
\expandafter\ifx\csname l@#1\endcsname\relax
\typeout{** WARNING: IEEEtran.bst: No hyphenation pattern has been}%
\typeout{** loaded for the language `#1'. Using the pattern for}%
\typeout{** the default language instead.}%
\else
\language=\csname l@#1\endcsname
\fi
#2}}
\providecommand{\BIBdecl}{\relax}
\BIBdecl

\bibitem{prabhavalkar2023end}
R.~Prabhavalkar, T.~Hori, T.~N. Sainath, R.~Schluter, and S.~Watanabe,
  ``End-to-end speech recognition: A survey,'' \emph{IEEE/ACM Transactions on
  Audio, Speech, and Language Processing}, vol.~32, pp. 325--351, 2023.

\bibitem{li2022recent}
J.~Li, ``Recent advances in end-to-end automatic speech recognition,''
  \emph{APSIPA Transactions on Signal and Information Processing}, vol.~11,
  no.~1, 2022.

\bibitem{ctc1}
A.~Graves, S.~Fern{\'a}ndez, F.~Gomez, and J.~Schmidhuber, ``Connectionist
  temporal classification: Labelling unsegmented sequence data with recurrent
  neural networks,'' in \emph{International Conference on Machine Learning},
  2006, pp. 369--376.

\bibitem{ctc2}
A.~Graves and N.~Jaitly, ``Towards end-to-end speech recognition with recurrent
  neural networks,'' in \emph{International Conference on Machine Learning},
  2014, pp. 1764--1772.

\bibitem{kriman2020quartznet}
S.~Kriman, S.~Beliaev, B.~Ginsburg, J.~Huang, O.~Kuchaiev, V.~Lavrukhin,
  R.~Leary, J.~Li, and Y.~Zhang, ``Quartznet: Deep automatic speech recognition
  with 1d time-channel separable convolutions,'' in \emph{IEEE International
  Conference on Acoustics, Speech, and Signal Processing}, 2020, pp.
  6124--6128.

\bibitem{rnnt1}
A.~Graves, ``Sequence transduction with recurrent neural networks,'' in
  \emph{International Conference on Machine Learning}, 2012.

\bibitem{zhang2020transformer}
Q.~Zhang, H.~Lu, H.~Sak, A.~Tripathi, E.~McDermott, S.~Koo, and S.~Kumar,
  ``Transformer transducer: A streamable speech recognition model with
  transformer encoders and rnn-t loss,'' in \emph{IEEE International Conference
  on Acoustics, Speech, and Signal Processing}, 2020, pp. 7829--7833.

\bibitem{han2020contextnet}
W.~Han, Z.~Zhang, Y.~Zhang, J.~Yu, C.-C. Chiu, J.~Qin, A.~Gulati, R.~Pang, and
  Y.~Wu, ``Contextnet: Improving convolutional neural networks for automatic
  speech recognition with global context,'' in \emph{Interspeech}, 2020, pp.
  3610--3614.

\bibitem{rnnt2}
K.~Rao, H.~Sak, and R.~Prabhavalkar, ``Exploring architectures, data and units
  for streaming end-to-end speech recognition with {RNN}-transducer,'' in
  \emph{” IEEE Workshop on Automatic Speech Recognition and Understanding},
  2017, pp. 193--199.

\bibitem{attention1}
W.~Chan, N.~Jaitly, Q.~Le, and O.~Vinyals, ``Listen, attend and spell: A neural
  network for large vocabulary conversational speech recognition,'' in
  \emph{IEEE International Conference on Acoustics, Speech, and Signal
  Processing}, 2016, pp. 4960--4964.

\bibitem{attention2}
J.~K. Chorowski, D.~Bahdanau, D.~Serdyuk, K.~Cho, and Y.~Bengio,
  ``Attention-based models for speech recognition,'' in \emph{Advances in
  Neural Information Processing Systems}, vol.~28, 2015, pp. 577--585.

\bibitem{karita2019comparative}
S.~Karita, N.~Chen, T.~Hayashi, T.~Hori, H.~Inaguma, Z.~Jiang, M.~Someki,
  N.~E.~Y. Soplin, R.~Yamamoto, X.~Wang \emph{et~al.}, ``A comparative study on
  transformer vs rnn in speech applications,'' in \emph{” IEEE Workshop on
  Automatic Speech Recognition and Understanding}, 2019, pp. 449--456.

\bibitem{guo2021recent}
P.~Guo, F.~Boyer, X.~Chang, T.~Hayashi, Y.~Higuchi, H.~Inaguma, N.~Kamo, C.~Li,
  D.~Garcia-Romero, J.~Shi \emph{et~al.}, ``Recent developments on espnet
  toolkit boosted by conformer,'' in \emph{IEEE International Conference on
  Acoustics, Speech, and Signal Processing}, 2021, pp. 5874--5878.

\bibitem{higuchi2020mask}
Y.~Higuchi, S.~Watanabe, N.~Chen, T.~Ogawa, and T.~Kobayashi, ``Mask ctc:
  Non-autoregressive end-to-end asr with ctc and mask predict,'' in
  \emph{Interspeech}, 2020, pp. 3655--3659.

\bibitem{chen2020non}
N.~Chen, S.~Watanabe, J.~Villalba, P.~{\.Z}elasko, and N.~Dehak,
  ``Non-autoregressive transformer for speech recognition,'' \emph{IEEE Signal
  Processing Letters}, vol.~28, pp. 121--125, 2020.

\bibitem{song2021non}
X.~Song, Z.~Wu, Y.~Huang, C.~Weng, D.~Su, and H.~Meng, ``Non-autoregressive
  transformer asr with ctc-enhanced decoder input,'' in \emph{IEEE
  International Conference on Acoustics, Speech, and Signal Processing}, 2021,
  pp. 5894--5898.

\bibitem{ctcsegmentation}
L.~K{\"u}rzinger, D.~Winkelbauer, L.~Li, T.~Watzel, and G.~Rigoll,
  ``Ctc-segmentation of large corpora for german end-to-end speech
  recognition,'' in \emph{Speech and Computer}, A.~Karpov and R.~Potapova,
  Eds.\hskip 1em plus 0.5em minus 0.4em\relax Cham: Springer International
  Publishing, 2020, pp. 267--278.

\bibitem{kuang22_interspeech}
F.~Kuang, L.~Guo, W.~Kang, L.~Lin, M.~Luo, Z.~Yao, and D.~Povey, ``{Pruned
  RNN-T for fast, memory-eﬀicient ASR training},'' in \emph{Interspeech},
  2022, pp. 2068--2072.

\bibitem{bahdanau2014neural}
D.~Bahdanau, K.~Cho, and Y.~Bengio, ``Neural machine translation by jointly
  learning to align and translate,'' in \emph{International Conference on
  Learning Representations}, 2015.

\bibitem{pmlr-v202-radford23a}
A.~Radford, J.~W. Kim, T.~Xu, G.~Brockman, C.~Mcleavey, and I.~Sutskever,
  ``Robust speech recognition via large-scale weak supervision,'' in
  \emph{International Conference on Machine Learning}, vol. 202.\hskip 1em plus
  0.5em minus 0.4em\relax PMLR, 2023, pp. 28\,492--28\,518.

\bibitem{watanabe2017hybrid}
S.~Watanabe, T.~Hori, S.~Kim, J.~R. Hershey, and T.~Hayashi, ``Hybrid
  ctc/attention architecture for end-to-end speech recognition,'' \emph{IEEE
  Journal of Selected Topics in Signal Processing}, vol.~11, no.~8, pp.
  1240--1253, 2017.

\bibitem{ghazvininejad2019mask}
M.~Ghazvininejad, O.~Levy, Y.~Liu, and L.~Zettlemoyer, ``Mask-predict: Parallel
  decoding of conditional masked language models,'' in \emph{Proc.
  EMNLP-IJCNLP}, 2019.

\bibitem{futami2022correction}
H.~Futami, H.~Inaguma, S.~Ueno, M.~Mimura, S.~Sakai, and T.~Kawahara,
  ``Non-autoregressive error correction for ctc-based asr with
  phone-conditioned masked lm,'' in \emph{Interspeech}, 2022, pp. 3889--3893.

\bibitem{ueno2018acoustic}
S.~Ueno, H.~Inaguma, M.~Mimura, and T.~Kawahara, ``Acoustic-to-word
  attention-based model complemented with character-level ctc-based model,'' in
  \emph{IEEE International Conference on Acoustics, Speech, and Signal
  Processing}, 2018, pp. 5804--5808.

\bibitem{nakatani2019improving}
T.~Nakatani, ``Improving transformer-based end-to-end speech recognition with
  connectionist temporal classification and language model integration,'' in
  \emph{Interspeech}, 2019.

\bibitem{peng2023reproducing}
Y.~Peng, J.~Tian, B.~Yan, D.~Berrebbi, X.~Chang, X.~Li, J.~Shi, S.~Arora,
  W.~Chen, R.~Sharma \emph{et~al.}, ``Reproducing whisper-style training using
  an open-source toolkit and publicly available data,'' in \emph{” IEEE
  Workshop on Automatic Speech Recognition and Understanding}, 2023, pp. 1--8.

\bibitem{peng2024owsm}
Y.~Peng, J.~Tian, W.~Chen, S.~Arora, B.~Yan, Y.~Sudo, S.~Muhammad, K.~Choi,
  J.~Shi, X.~Chang \emph{et~al.}, ``{OWSM v3.1}: Better and faster open
  whisper-style speech models based on {E-Branchformer},'' in
  \emph{Interspeech}, 2024, pp. 352--356.

\bibitem{zhang2022wenetspeech}
B.~Zhang, H.~Lv, P.~Guo, Q.~Shao, C.~Yang, L.~Xie, X.~Xu, H.~Bu, X.~Chen,
  C.~Zeng \emph{et~al.}, ``Wenetspeech: A 10000+ hours multi-domain mandarin
  corpus for speech recognition,'' in \emph{IEEE International Conference on
  Acoustics, Speech, and Signal Processing}, 2022, pp. 6182--6186.

\bibitem{fujimotoreazonspeech}
Y.~Yin, D.~Mori, and S.~Fujimoto, ``Reazonspeech: A free and massive corpus for
  japanese asr,'' in \emph{The Association for Natural Language Processing},
  2023, pp. 1134--1139.

\bibitem{sainath2019two}
T.~N. Sainath, R.~Pang, D.~Rybach, Y.~He, R.~Prabhavalkar, W.~Li, M.~Visontai,
  Q.~Liang, T.~Strohman, Y.~Wu \emph{et~al.}, ``Two-pass end-to-end speech
  recognition,'' in \emph{Interspeech}, 2019, pp. 2713--2777.

\bibitem{hu2020deliberation}
K.~Hu, T.~N. Sainath, R.~Pang, and R.~Prabhavalkar, ``Deliberation model based
  two-pass end-to-end speech recognition,'' in \emph{IEEE International
  Conference on Acoustics, Speech, and Signal Processing}.\hskip 1em plus 0.5em
  minus 0.4em\relax IEEE, 2020, pp. 7799--7803.

\bibitem{hu2021transformer}
K.~Hu, R.~Pang, T.~N. Sainath, and T.~Strohman, ``Transformer based
  deliberation for two-pass speech recognition,'' in \emph{IEEE Spoken Language
  Technology Workshop}, 2021, pp. 68--74.

\bibitem{tian2022hybrid}
Z.~Tian, J.~Yi, J.~Tao, S.~Zhang, and Z.~Wen, ``Hybrid autoregressive and
  non-autoregressive transformer models for speech recognition,'' \emph{IEEE
  Signal Processing Letters}, vol.~29, pp. 762--766, 2022.

\bibitem{wang2022deliberation}
W.~Wang, K.~Hu, and T.~N. Sainath, ``Deliberation of streaming rnn-transducer
  by non-autoregressive decoding,'' in \emph{IEEE International Conference on
  Acoustics, Speech, and Signal Processing}, 2022, pp. 7452--7456.

\bibitem{yao2021wenet}
Z.~Yao, D.~Wu, X.~Wang, B.~Zhang, F.~Yu, C.~Yang, Z.~Peng, X.~Chen, L.~Xie, and
  X.~Lei, ``Wenet: Production oriented streaming and non-streaming end-to-end
  speech recognition toolkit,'' in \emph{Interspeech}, 2021, pp. 4045--4058.

\bibitem{Narayanan2020CascadedEF}
A.~Narayanan, T.~N. Sainath, R.~Pang, J.~Yu, C.-C. Chiu, R.~Prabhavalkar,
  E.~Variani, and T.~Strohman, ``Cascaded encoders for unifying streaming and
  non-streaming asr,'' in \emph{IEEE International Conference on Acoustics,
  Speech, and Signal Processing}, 2020, pp. 5629--5633.

\bibitem{mahadeokar2022streaming}
J.~Mahadeokar, Y.~Shi, K.~Li, D.~Le, J.~Zhu, V.~Chandra, O.~Kalinli, and M.~L.
  Seltzer, ``Streaming parallel transducer beam search with fast-slow cascaded
  encoders,'' in \emph{Interspeech}, 2022, pp. 2083--2087.

\bibitem{li2022improving}
K.~Li, J.~Mahadeokar, J.~Guo, Y.~Shi, G.~Keren, O.~Kalinli, M.~L. Seltzer, and
  D.~Le, ``Improving fast-slow encoder based transducer with streaming
  deliberation,'' \emph{IEEE International Conference on Acoustics, Speech, and
  Signal Processing}, pp. 1--5, 2023.

\bibitem{yu2021dualmode}
J.~Yu, W.~Han, A.~Gulati, C.-C. Chiu, B.~Li, T.~N. Sainath, Y.~Wu, and R.~Pang,
  ``Dual-mode {ASR}: Unify and improve streaming {ASR} with full-context
  modeling,'' in \emph{International Conference on Learning Representations},
  2021.

\bibitem{moritz2021dual}
N.~Moritz, T.~Hori, and J.~L. Roux, ``Dual causal/non-causal self-attention for
  streaming end-to-end speech recognition,'' in \emph{Interspeech}, 2021, pp.
  1822--1826.

\bibitem{weninger2022conformer}
F.~Weninger, M.~Gaudesi, M.~A. Haidar, N.~Ferri, J.~Andr{\'e}s-Ferrer, and
  P.~Zhan, ``Conformer with dual-mode chunked attention for joint online and
  offline asr,'' in \emph{Interspeech}, 2022, pp. 2053--2057.

\bibitem{hinton2012deep}
G.~Hinton, L.~Deng, D.~Yu, G.~E. Dahl, A.-r. Mohamed, N.~Jaitly, A.~Senior,
  V.~Vanhoucke, P.~Nguyen, T.~N. Sainath \emph{et~al.}, ``Deep neural networks
  for acoustic modeling in speech recognition: The shared views of four
  research groups,'' \emph{IEEE Signal processing magazine}, vol.~29, no.~6,
  pp. 82--97, 2012.

\bibitem{seide11_interspeech}
F.~Seide, G.~Li, and D.~Yu, ``Conversational speech transcription using
  context-dependent deep neural networks,'' in \emph{Interspeech}, 2011, pp.
  437--440.

\bibitem{XU2011802}
H.~Xu, D.~Povey, L.~Mangu, and J.~Zhu, ``Minimum bayes risk decoding and system
  combination based on a recursion for edit distance,'' \emph{Computer Speech
  \& Language}, vol.~25, no.~4, pp. 802--828, 2011.

\bibitem{6638967}
P.~Swietojanski, A.~Ghoshal, and S.~Renals, ``Revisiting hybrid and gmm-hmm
  system combination techniques,'' in \emph{IEEE International Conference on
  Acoustics, Speech, and Signal Processing}, 2013, pp. 6744--6748.

\bibitem{wang15k_interspeech}
H.~Wang, A.~Ragni, M.~J.~F. Gales, K.~M. Knill, P.~C. Woodland, and C.~Zhang,
  ``Joint decoding of tandem and hybrid systems for improved keyword spotting
  on low resource languages,'' in \emph{Interspeech}, 2015, pp. 3660--3664.

\bibitem{7472764}
J.~Yang, C.~Zhang, A.~Ragni, M.~J.~F. Gales, and P.~C. Woodland, ``System
  combination with log-linear models,'' in \emph{IEEE International Conference
  on Acoustics, Speech, and Signal Processing}, 2016, pp. 5675--5679.

\bibitem{wong20_interspeech}
J.~H. Wong, Y.~Gaur, R.~Zhao, L.~Lu, E.~Sun, J.~Li, and Y.~Gong, ``Combination
  of end-to-end and hybrid models for speech recognition,'' in
  \emph{Interspeech}, 2020, pp. 1783--1787.

\bibitem{alumae21_interspeech}
T.~Alumäe and J.~Kong, ``Combining hybrid and end-to-end approaches for the
  openasr20 challenge,'' in \emph{Interspeech}, 2021, pp. 4349--4353.

\bibitem{9747144}
G.~Ye, V.~Mazalov, J.~Li, and Y.~Gong, ``Have best of both worlds: Two-pass
  hybrid and e2e cascading framework for speech recognition,'' in \emph{IEEE
  International Conference on Acoustics, Speech, and Signal Processing}, 2022,
  pp. 7432--7436.

\bibitem{LI202312}
Q.~Li, C.~Zhang, and P.~C. Woodland, ``Combining hybrid dnn-hmm asr systems
  with attention-based models using lattice rescoring,'' \emph{Speech
  Communication}, vol. 147, pp. 12--21, 2023.

\bibitem{yan2022ctc}
B.~Yan, S.~Dalmia, Y.~Higuchi, G.~Neubig, F.~Metze, A.~W. Black, and
  S.~Watanabe, ``Ctc alignments improve autoregressive translation,'' in
  \emph{Proceedings of the 17th Conference of the European Chapter of the
  Association for Computational Linguistics}, 2022, pp. 1623--1639.

\bibitem{tsunoo23_interspeech}
E.~Tsunoo, H.~Futami, Y.~Kashiwagi, S.~Arora, and S.~Watanabe, ``{Integration
  of Frame- and Label-synchronous Beam Search for Streaming Encoder-decoder
  Speech Recognition},'' in \emph{Interspeech}, 2023, pp. 1369--1373.

\bibitem{sudo23c_interspeech}
Y.~Sudo, S.~Muhammad, Y.~Peng, and S.~Watanabe, ``{Time-synchronous one-pass
  Beam Search for Parallel Online and Offline Transducers with Dynamic Block
  Training},'' in \emph{Interspeech}, 2023, pp. 4479--4483.

\bibitem{sudo20224d}
Y.~Sudo, S.~Muhammad, B.~Yan, J.~Shi, and S.~Watanabe, ``4{D} {ASR}: Joint
  modeling of {CTC}, attention, transducer, and mask-predict decoders,'' in
  \emph{Interspeech}, 2023, pp. 3312--3316.

\bibitem{gulati2020conformer}
A.~Gulati, C.~Chiu, J.~Qin, J.~Yu, N.~Parmar, R.~Pang, S.~Wang, W.~Han, Y.~Wu,
  Y.~Zhang, and Z.~Zhang, ``Conformer: Convolution-augmented transformer for
  speech recognition,'' in \emph{Interspeech}, 2020, pp. 5036--5040.

\bibitem{ba2016layer}
J.~L. Ba, J.~R. Kiros, and G.~E. Hinton, ``Layer normalization,''
  \emph{Advances in Neural Information Processing Systems}, 2016.

\bibitem{7780459}
K.~He, X.~Zhang, S.~Ren, and J.~Sun, ``Deep residual learning for image
  recognition,'' in \emph{The IEEE/CVF Conference on Computer Vision and
  Pattern Recognition}, 2016, pp. 770--778.

\bibitem{ma2019monotonic}
X.~Ma, J.~Pino, J.~Cross, L.~Puzon, and J.~Gu, ``Monotonic multihead
  attention,'' in \emph{International Conference on Learning Representations},
  2020, pp. 1--15.

\bibitem{inaguma20b_interspeech}
H.~Inaguma, M.~Mimura, and T.~Kawahara, ``Enhancing monotonic multihead
  attention for streaming asr,'' in \emph{Interspeech}, 2020, pp. 2137--2141.

\bibitem{Lin2019AdaptiveAT}
X.~Lin, H.~S. Baweja, G.~A. Kantor, and D.~Held, ``Adaptive auxiliary task
  weighting for reinforcement learning,'' in \emph{Advances in Neural
  Information Processing Systems}, 2019.

\bibitem{moritz2019triggered}
N.~Moritz, T.~Hori, and J.~Le~Roux, ``Triggered attention for end-to-end speech
  recognition,'' in \emph{IEEE International Conference on Acoustics, Speech,
  and Signal Processing}, 2019, pp. 5666--5670.

\bibitem{specaug}
D.~S. Park, W.~Chan, Y.~Zhang, C.~Chiu, B.~Zoph, E.~D. Cubuk, and Q.~V. Le,
  ``{SpecAugment}: A simple data augmentation method for automatic speech
  recognition,'' in \emph{Interspeech}, 2019, pp. 2613--2617.

\bibitem{kingma2014adam}
D.~P. Kingma and J.~Ba, ``Adam: A method for stochastic optimization,''
  \emph{CoRR}, vol. abs/1412.6980, 2015.

\bibitem{espnet}
S.~Watanabe, T.~Hori, S.~Karita, T.~Hayashi, J.~Nishitoba, Y.~Unno, N.~{Enrique
  Yalta Soplin}, J.~Heymann, M.~Wiesner, N.~Chen, A.~Renduchintala, and
  T.~Ochiai, ``{ESPnet}: End-to-end speech processing toolkit,'' in
  \emph{Interspeech}, 2018, pp. 2207--2211.

\bibitem{panayotov2015librispeech}
V.~Panayotov, G.~Chen, D.~Povey, and S.~Khudanpur, ``Librispeech: an asr corpus
  based on public domain audio books,'' in \emph{IEEE International Conference
  on Acoustics, Speech, and Signal Processing}, 2015, pp. 5206--5210.

\bibitem{csj}
K.~Maekawa, ``Corpus of spontaneous {Japanese}: Its design and evaluation,'' in
  \emph{ISCA/IEEE Workshop on Spontaneous Speech Processing and Recognition},
  2003.

\bibitem{KUREMATSU1990357}
A.~Kurematsu, K.~Takeda, Y.~Sagisaka, S.~Katagiri, H.~Kuwabara, and K.~Shikano,
  ``Atr japanese speech database as a tool of speech recognition and
  synthesis,'' \emph{Speech Communication}, vol.~9, no.~4, pp. 357--363, 1990.

\bibitem{fiscus1997post}
J.~G. Fiscus, ``A post-processing system to yield reduced word error rates:
  Recognizer output voting error reduction (rover),'' in \emph{” IEEE
  Workshop on Automatic Speech Recognition and Understanding}, 1997, pp.
  347--354.

\end{thebibliography}
